\documentclass{pasj01}

\RequirePackage{mathpazo}
\RequirePackage[T1]{fontenc}

\def\commentb{$^\dagger$}

\newcounter{author}
\def\authorcount#1#2{\refstepcounter{author}\label{#1}
                     \altaffiltext{\ref{#1}}{#2}}

\begin{document}

\SetRunningHead{A.Imada et al.}{}

\Received{2017/06/01}
\Accepted{2017/11/15}

\title{OAO/MITSuME Photometry of Dwarf Novae. \\II. HV Virginis and OT
  J012059.6+325545}

\author{Akira~\textsc{Imada},\altaffilmark{\ref{affil:hamburg}}$^,$\altaffilmark{\ref{affil:kwasan}*}
  Keisuke~\textsc{Isogai},\altaffilmark{\ref{affil:kyoto}}
  Takahiro~\textsc{Araki},\altaffilmark{\ref{affil:kagoshima}}
  Shunsuke~\textsc{Tanada},\altaffilmark{\ref{affil:kagoshima}}
  Kenshi~\textsc{Yanagisawa},\altaffilmark{\ref{affil:oao}}
  Nobuyuki~\textsc{Kawai}\altaffilmark{\ref{affil:ttec}}
}

\authorcount{affil:hamburg}{Hamburger Sternwarte, Universit\"at Hamburg, Gojenbergsweg 112, D-21029 Hamburg, Germany}
\email{$^*$a\_imada@kusastro.kyoto-u.ac.jp}

\authorcount{affil:kwasan}{
  Kwasan and Hida Observatories, Kyoto University, Yamashina, Kyoto 607-8471, Japan}

\authorcount{affil:kyoto}{
  Department of Astronomy, Kyoto University, Kyoto 606-8502, Japan}

\authorcount{affil:kagoshima}{
  Faculty of Science, Kagoshima University, 1-21-30 Korimoto, Kagoshima, Kagoshima 890-0065, Japan}

\authorcount{affil:oao}{
  Okayama Astrophysical Observatory, National Astronomical Observatory of Japan, Asakuchi, Okayama 719-0232, Japan}

\authorcount{affil:ttec}{
Department of Physics, Tokyo Institute of Technology, Ookayama 2-12-1, Meguro-ku, Tokyo 152-8551, Japan}


 
\maketitle

\begin{abstract}

We report on multicolor photometry of WZ Sge-type dwarf novae, HV Vir and OT J012059.6+325545 during superoutbursts. These systems show early superhumps with the mean periods of 0.057093(45) d for HV Vir and 0.057147(15) d for OT
J012059.6+325545, respectively. The observed early superhumps showed a
common feature that the brightness minima correspond to the bluest
peaks in color variations, which may be a ubiquitous phenomenon among
early superhumps of WZ Sge-type dwarf novae. We confirmed that amplitudes of early superhumps depend on wavelength: amplitudes with longer bandpass filters show larger values. This indicates that the light source of early superhumps is generated at the outer region of the vertically-extended accretion disk. On the other hand, amplitudes of ordinary superhumps are likely to be independent of wavelength. This implies that the superhump light source is geometrically thin. We also examined color variations of ordinary superhumps and found that the bluest peaks in $g'-I_{\rm c}$ tend to coincide with the brightness minima, particularily in stage B superhumps. This may reflect that the pressure effect plays a dominant role during stage B superhumps.

\end{abstract}

\section{Introduction}

Cataclysmic variables (CVs) are close binary systems that consist of
a primary white dwarf and a secondary star. The secondary star fills 
its Roche lobe, transferring matter into the primary Roche lobe 
via the inner Lagrangian point (L1). If the magnetic field of the white dwarf is weak, an accretion disk is formed around the white dwarf (for a review, see, e.g., \cite{war95book}; \cite{hel01book}).

Dwarf novae are a subclass of CVs, exhibiting outbursts with recurrent
time scales of days to years and amplitudes of 2$-$8 mag (for a
review, see, e.g., \cite{osa96review}; \cite{las01DIDNXT}; \cite{VSNET}). The accretion disk plays a major role on the overall behavior of dwarf novae, which is well explained by the thermal instability model of the accretion disk
(\cite{mey81DNoutburst}; \cite{sma84DI}).

Dwarf novae are further divided into several subclasses according to
their light curves (\cite{VSNET}; \cite{pdot}). One of the subclasses is SU UMa-type dwarf novae. These show two types of outbursts. One is normal outburst, whose duration is a few days, and the other is superoutburst, whose duration is typically 2 weeks. During the superoutburst, modulations with a rapid rise and slow decline, called (positive) superhumps, are visible. The period of the superhump ($P_{\rm sh}$) is slightly longer than the orbital period of the system ($P_{\rm orb}$).\footnote{Some dwarf novae show negative superhumps, whose period is slightly shorter than $P_{\rm orb}$. Throughout this paper, the term "superhump" means positive superhump.} Long-term light curves of SU UMa-type dwarf novae are well explained by the combination of the thermal and tidal instability model (\cite{osa89suuma}; \cite{osa13v1504cygKepler}), and the basic properties of superhumps are understood by tidal dissipation of the eccentrically-deformed precessing accretion disk
(\cite{whi88tidal}; \cite{hir90SHexcess}).

Is has been widely recognized that some SU UMa-type dwarf novae show
unusual behavior in their short and long-term light curves. WZ
Sge-type dwarf novae, one of the subclasses of SU UMa-type dwarf novae, show unique properties during superoutburst and quiescence \citep{kat15wzsge}. Their main characteristic features are that (1) their quiescence is unusually long, sometimes exceeding a decade (\cite{nog97alcom};
\cite{kat01hvvir}) (2) the amplitudes of the superoutburst exceed 6 mag (\cite{how95TOAD}; \cite{ish01rzleo}), (3) an early stage of the
superoutburst contains double-peaked modulations called early
superhumps (\cite{osa02wzsgehump}; \cite{kat02wzsgeESH})\footnote{Throughout this paper, a period of early superhumps is designated as $P_{\rm Esh}$.}, (4)
they show rebrightening(s) after the termination of the main
superoutburst (\cite{pat02wzsge}; \cite{kat04egcnc}), and (5) they are absent of normal outbursts (For a comprehensive review, see \citet{kat15wzsge}).
 Recent optical
transient surveys, such as ASAS \citep{ASASSN}, MASTER \citep{MASTER0}, and CRTS \citep{crts} have revealed the presence of extensive numbers of candidates for WZ Sge-type dwarf novae. However, most of photometry was performed without filters. This means that we have less information on color and temperature variations in the accretion disk of WZ Sge-type dwarf novae.

Based on extensive photometry during superoutbursts, T.Kato
and his colleagues have established the "textbook'' behavior of superhump
period changes \citep{pdot}. According to \citet{pdot}, the evolution of the
superhump period is composed of three stages: an early stage with a longer and constant superhump period (stage A), a middle stage with positive period derivatives $P_{\rm dot}$ = ${\dot P}/P$ (stage B), and a late stage with a shorter and constant period (stage C). Although the majority of SU UMa-type dwarf novae follow the trend described here, some systems show an unusual period evolution of superhumps \citep{pdot4}. In order to understand the diversity of superhump period changes, multicolor photometry may be an attractive approach. Indeed, simultaneous optical/near-infrared
observations have revealed the presence of the warm accretion disk
after the termination of a superoutburst \citep{uem08j1021}, and difference in maximum timings between magnitudes and colors of superhumps (\cite{sch80vwhyi}; \cite{has85ektra}; \cite{iso15ezlyn}).

Here we report on optical-near infrared photometry of HV Vir
and optical photometry of OT J012059.6+325545 during superoutbursts by using instruments of Okayama Astrophysical Observatory\footnote{http://www.oao.nao.ac.jp}. In the following
subsection, we outline the background of these stars.

\subsection{HV Virginis}

HV Vir is one of well-known WZ Sge-type dwarf novae with magnitudes
ranging from 11.5 to 19 \citep{RKcat}. The system underwent
outbursts in 1929, 1939, 1970, 1981, 1992, 2008, and 2016
\citep{pdot9}. Superhumps and early superhumps were
detected during the 1992 superoutburst, from which HV Vir has been
recognized as a member of WZ Sge-type dwarf novae (\cite{bar92hvvir};
\cite{lei94hvvir}; \cite{how95TOAD}). \citet{kat01hvvir} derived the
mean early superhump period to be $P_{\rm Esh}$ = 0.057085(23) d.
A photometric campaign of the 2002 superoutburst was
carried out by \citet{ish03hvvir}, who obtained $P_{\rm sh}$ = 0.058203(1) d for
superhumps and $P_{\rm Esh}$ = 0.0569(1) d for early superhumps,
respectively. \citet{ish03hvvir} also derived the rate of the period
change during stage B to be $P_{\rm dot}$ = 7.8(7) $\times$
10$^{-5}$. \citet{pdot} obtained $P_{\rm dot}$ = 7.1(1.9) $\times$
10$^{-5}$ for the 2008 superoutburst. \citet{pat03suumas} performed quiescent photometry and derived the orbital period to be $P_{\rm orb}$ = 0.057069(6) d. \citet{pat03suumas} also noted the presence of a signal at 11.20 or 12.21 c/d (P=0.0893 or 0.0819 d), the origin of which is unknown. Similar periodicities were also reported by \citet{wou12SDSSCRTSCVs}, who observed HV Vir in quiescence in 2010 April. An estimation of the temperature of the white dwarf yielded a cool $T_{\rm eff}$ of 13300$\pm$800 K \citep{szk02egcnchvvirHST}. A distance to the object is estimated as 480$\pm$70 pc \citep{szk02egcnchvvirHST}, 460(+530,-180) pc \citep{tho03CVdistance}, and 300 pc \citep{pat11CVdistance}, respectively.

Here we report on analyses and results of simultaneous optical-near infrared photometry during the 2008 superoutburst of HV Vir. On 2008 February 14.601 (JD 2454511.101), Rod Stubbings detected an eruption of HV Vir with a visual magnitude of 11.5 ([vsnet-alert 9912])\footnote{http://ooruri.kusastro.kyoto-u.ac.jp/pipermail/vsnet-alert/ . For a review of VSNET, see \citet{VSNET}}. P. Schmeer reported that HV Vir was fainter than 13.7 on 2008 February 14.139 (JD 2454510.639) ([vsnet-alert 9915]). In combination with these reports, we succeeded in observing from the very onset of the outburst.

\subsection{OT J012059.6+325545}

OT J012059.6+325545 (hereafter, J0120) was discovered by K. Itagaki as
a transient object on 2010 November 30.50663 (JD 2455531.00663) with a
magnitude of 12.3 ([vsnet-alert 12431]). According to the SDSS DR9
database, quiescent magnitudes are $u$ = 20.359, $g$ = 20.088, $r$ =
20.237, $i$ = 20.518, and $z$ = 21.216 \citep{sdss9}. The object has
no 2MASS counterpart within 5 arcsec. A distance to the object is
estimated as 355 pc \citep{kat12DNSDSS}. After the discovery of the outburst, VSNET conducted a photometric campaign and detected early superhumps with a mean period of $P_{\rm Esh}$ = 0.057155(5) d, from which J0120 was qualified as a WZ Sge-type dwarf nova \citep{pdot3}. \citet{nak13j0120} reported on simultaneous $g'$, $R_{\rm c}$, and $i'$ photometry during the 2010$-$2011 superoutburst. Using a model developed by \citet{uem12ESHrecon}, they calculated disk structures during the early superhump stage. 

In this paper, we focus mainly on details of color and amplitude variations of early and ordinary superhumps which \citet{nak13j0120} did not deal with. We succeeded in detecting stage A superhumps in our data. By using a new method of estimating the mass ratio \citep{kat13qfromstageA}, we derived the mass ratio of the system for the first time.

\section{Observations}

\begin{table}
\caption{Log of optical observations of HV Vir using MITSuME.}
\begin{center}
\begin{tabular}{cccc}
\hline\hline
Date & JD(start)$^*$ & JD(end)$^*$ & N$^{\dagger}$ \\
\hline
2008 February 15 & 4512.1196 & 4512.3684 & 360 \\ 
2008 February 16 & 4513.1156 & 4513.3681 & 220 \\ 
2008 February 17 & 4514.1111 & 4514.3638 & 430 \\
2008 February 18 & 4515.1103 & 4515.3658 & 387 \\
2008 February 19 & 4516.0684 & 4516.3685 & 542 \\
2008 February 20 & 4517.1018 & 4517.3707 & 326 \\
2008 February 21 & 4518.1024 & 4518.3470 & 218 \\
2008 February 22 & 4519.2506 & 4519.3710 & 33 \\ 
2008 February 23 & 4520.2479 & 4520.3149 & 44 \\ 
2008 February 26 & 4523.0950 & 4523.3655 & 232 \\ 
2008 February 27 & 4524.0904 & 4524.3600 & 274 \\ 
2008 February 28 & 4525.0821 & 4525.3472 & 264 \\ 
2008 February 29 & 4526.0988 & 4526.2114 & 91 \\ 
2008 March 1 & 4527.0560 & 4527.3569 & 216 \\ 
2008 March 2 & 4528.0954 & 4528.2019 & 105 \\ 
2008 March 3 & 4529.2393 & 4529.3154 & 3 \\ 
2008 March 4 & 4530.0827 & 4530.3550 & 132 \\ 
2008 March 5 & 4531.0996 & 4531.3552 & 254 \\ 
2008 March 6 & 4532.2276 & 4532.3553 & 87 \\ 
2008 March 7 & 4533.0977 & 4533.3540 & 200 \\ 
2008 March 8 & 4534.0887 & 4534.3519 & 131 \\ 
2008 March 11 & 4537.1575 & 4537.3261 & 130 \\ 
2008 March 12 & 4538.1493 & 4538.3242 & 102 \\ 
2008 March 14 & 4540.2031 & 4540.2988 & 71 \\
2008 March 15 & 4541.2082 & 4541.2899 & 55 \\ 
2008 March 16 & 4542.2523 & 4542.2772 & 10 \\ 
2008 March 17 & 4543.2471 & 4543.2873 & 3 \\ 
2008 March 31 & 4557.1926 & 4557.2003 & 3 \\ 
\hline
\multicolumn{4}{l}{$^*$JD$-$2450000 $^{\dagger}$ Number of exposure.}
\end{tabular}
\end{center}
\label{hvvir_log}
\end{table}

\begin{table}
\caption{Log of near-infrared observations of HV Vir using ISLE.}
\begin{center}
\begin{tabular}{ccccc}
\hline\hline
Date & JD(start)$^*$ & JD(end)$^*$ & N$^\dagger$ & band$^\S$\\
\hline
2008 February 14 & 4511.2348 & 4511.3461 & 230 & $H$ \\
2008 February 15 & 4512.1723 & 4512.2501 & 123 & $H$ \\
2008 February 15 & 4512.2509 & 4512.3666 & 185 & $K_{\rm s}$ \\
2008 February 16 & 4513.1748 & 4513.3858 & 167 &  $K_{\rm s}$ \\
2008 February 17 & 4514.1632 & 4514.3231 & 178 &  $K_{\rm s}$ \\
2008 February 18 & 4515.1856 & 4515.3538 & 179 &  $K_{\rm s}$ \\
2008 February 19 & 4516.1788 & 4516.3593 & 337 &  $K_{\rm s}$ \\
\hline
\multicolumn{4}{l}{$^*$JD$-$2450000 $^\dagger$ Number of exposure.} \\
\multicolumn{4}{l}{$^\S$ $H$: $H$ band. $K_{\rm s}$: $K_{\rm s}$ band.} \\
\end{tabular}
\end{center}
\label{hvvir_log_isle}
\end{table}

\begin{table}
\caption{Log of observations for J0120.}
\begin{center}
\begin{tabular}{cccc}
\hline\hline
Date & JD(start)$^{*}$ & JD(end)$^{*}$ & N$^{\dagger}$ \\
\hline
2010 December 3 & 5533.8857 & 5534.0883 & 445\\
2010 December 4 & 5534.9216 & 5535.1766 & 319\\
2010 December 5 & 5535.8853 & 5536.0879 & 446\\
2010 December 9 & 5539.9196 & 5540.0256 & 234\\
2010 December 10 & 5540.9302 & 5541.0535 & 223\\
2010 December 11 & 5541.8968 & 5542.0229 & 243\\
2010 December 12 & 5542.9288 & 5543.1383 & 24\\
2010 December 14 & 5544.9190 & 5545.0654 & 183\\
2010 December 15 & 5545.8884 & 5546.0053 & 18\\
2010 December 17 & 5547.9057 & 5547.9982 & 9\\
2010 December 18 & 5548.9637 & 5549.1209 & 174\\
2010 December 19 & 5549.8947 & 5549.9992 & 206\\
2010 December 20 & 5550.9093 & 5551.1697 & 443\\
2010 December 22 & 5552.8977 & 5553.1289 & 243\\
2010 December 23 & 5553.8743 & 5554.1251 & 341\\
2010 December 24 & 5554.9928 & 5555.1228 & 2\\
2010 December 25 & 5555.8750 & 5556.1277 & 294\\
2010 December 26 & 5556.9594 & 5557.1822 & 5\\
2010 December 27 & 5557.8760 & 5558.0998 & 120\\
2010 December 28 & 5558.9161 & 5558.9577 & 56\\
2010 December 29 & 5559.8769 & 5560.1004 & 400\\
2010 December 30 & 5560.8774 & 5561.0367 & 229\\
2010 December 31 & 5561.8778 & 5562.0846 & 7\\
2011 January 1 & 5562.8784 & 5562.9930 & 204\\
2011 January 2 & 5563.8792 & 5564.1363 & 142\\
2011 January 4 & 5565.8802 & 5565.8846 & 11\\
\hline
\multicolumn{4}{l}{$^{*}$JD$-$2450000} \\
\multicolumn{4}{l}{$^{\dagger}$Number of exposure.} \\
\end{tabular}
\end{center}
\label{j0120_log}
\end{table}

Time-resolved photometry were performed from 2008 February 14 to March
31 (JD 2454511$-$2454557, for HV Vir) and from 2010 December 3 to
2011 January 4 (JD 2455533$-$2455565 for J0120), respectively. We used OAO/MITSuME 50cm-telescope and OAO/ISLE 188-cm telescope. We obtained $g'$, $R_{\rm c}$, $I_{\rm c}$, $H$, and $K_{\rm s}$ bands for HV Vir, while we obtained  $g'$, $R_{\rm c}$, and $I_{\rm c}$ bands for J0120. OAO/MITSuME 50cm-telescope is a robotic telescope that can obtain $g'$, $R_{\rm c}$, and $I_{\rm c}$ simultaneously by using two dichroic mirrors and three CCD cameras, which enables us to study color variations of variable stars (For a detail description, see \cite{3me}). ISLE is an instrument that obtains $J$, $H$, and $K_{\rm s}$ images and medium-resolution spectra, attached with the Cassegrain focus of the OAO 188-cm
telescope \citep{ISLE1}. Table \ref{hvvir_log}, \ref{hvvir_log_isle}, and
\ref{j0120_log} show the journals of observations. All data were
obtained with 30$-$sec exposures.

After de-biasing, dark-subtracting and flat-fielding the images
with the standard procedure, the data were analyzed with aperture
photometry using IRAF/daophot.\footnote{IRAF (Image Reduction and
  Analysis Facility) is distributed by the National Optical Astronomy
  Observatories, which is operated by the Association of Universities
  for Research in Astronomy, Inc., under cooperative agreement with
  National Science Foundation.} In order to study color variations of $g'-K_{\rm s}$, we interpolated the ISLE magnitudes to the times of MITSuME observations, because there were slight differences of mid-exposure time between the data of MITSuME and ISLE. We derived
magnitudes and colors with differential photometry using SDSS J132054.81+015207.4 (g' = 14.286(3), r' = 13.938(4), i' = 13.838(4)) for HV Vir and 
SDSS J012048.61+325718.1 (g' = 12.779(1), r' = 12.522(1), i' =
12.624(2)) for J0120, respectively. The constancy of these
comparison stars is checked by nearby stars in the same field. Because
the exact magnitudes of $R_{\rm c}$ and $I_{\rm c}$ of the comparison
star are unknown, we converted the SDSS magnitudes to $R_{\rm c}$
and $I_{\rm c}$ given by \citet{2002AJ....123.2121S} as follows:

\begin{equation}
V = g' - 0.55(g' - r') - 0.03
\end{equation}
\begin{equation}
V - R = 0.59(g' - r')+ 0.11
\end{equation}
\begin{equation}
R - I = 1.00(r' - i') + 0.21
\end{equation}

Using the above equations and 2MASS archives, we adopt $g'$ = 14.286(3), 
$R_{\rm c}$ = 13.749(5), $I_{\rm c}$ = 13.439(5), $H$ = 12.716(25), and 
$K_{\rm s}$ = 12.686(27) as the magnitudes of the comparison
star of HV Vir. As for the magnitudes of the comparison star of J0120,
we adopt $g'$ = 12.779(1), $R_{\rm c}$ = 12.346(1) and $I_{\rm c}$ =
12.238(2). Barycentric correction was made before the following analyses.

\section{Results of HV Vir}

\subsection{overall light curves}

\begin{figure}[htb]
\begin{center}
\FigureFile(80mm,80mm){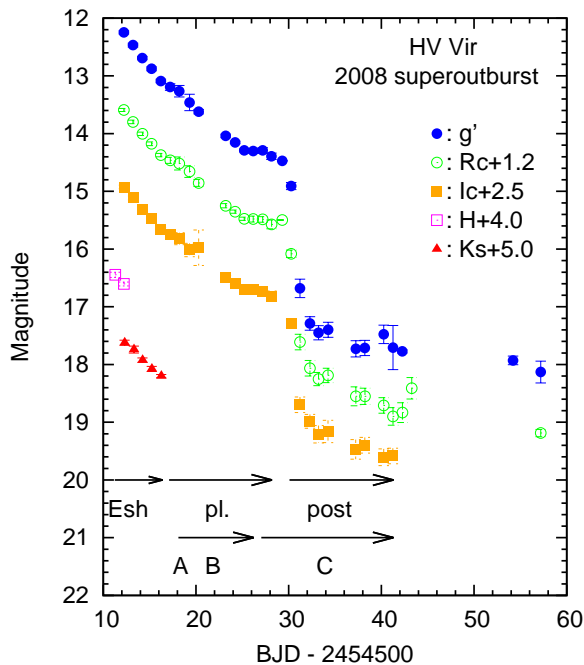}
\end{center}
\caption{Light curves of HV Vir during the 2008
  superoutburst. The obtained data except for $g'$ band are shifted
  for display purposes. As noted above, a visual magnitude was fainter
  than 13.7 on JD 2454510.639, after which the outburst was discovered
  on JD 2454511.101 with a visual magnitude of 11.5. After the termination of
  the plateau stage, a long fading tail was observed. No rebrightening
  was recorded during our run.}
\label{lc_hvvir}
\end{figure}

\begin{figure}[htb]
\begin{center}
\FigureFile(80mm,80mm){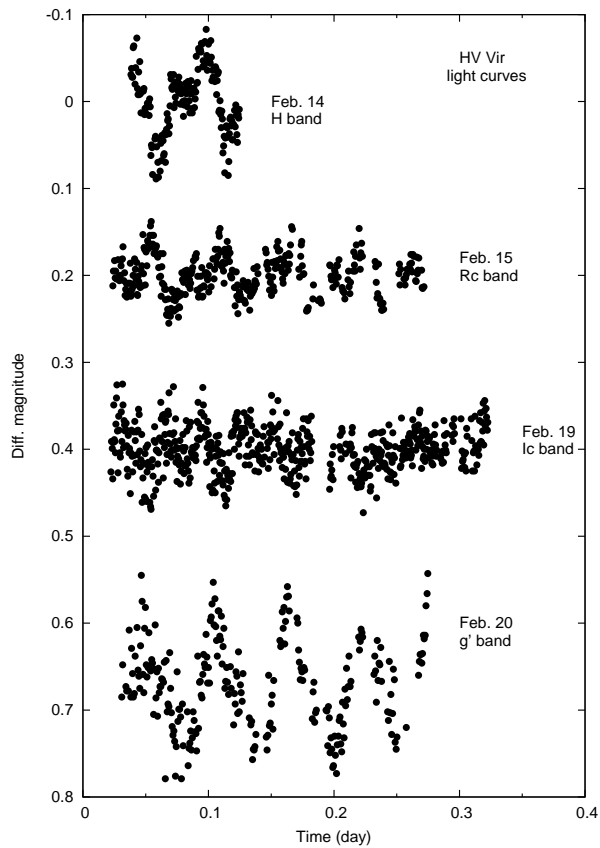}
\end{center}
\caption{Representative light curves of HV Vir. Early superhumps are visible on February 14, 15, and 19, while ordinary superhumps are seen on February 20. The light curves are arbitrarily shifted for better visualization.}
\label{rep_hvvir}
\end{figure}

\begin{figure}[htb]
\begin{center}
\FigureFile(80mm,80mm){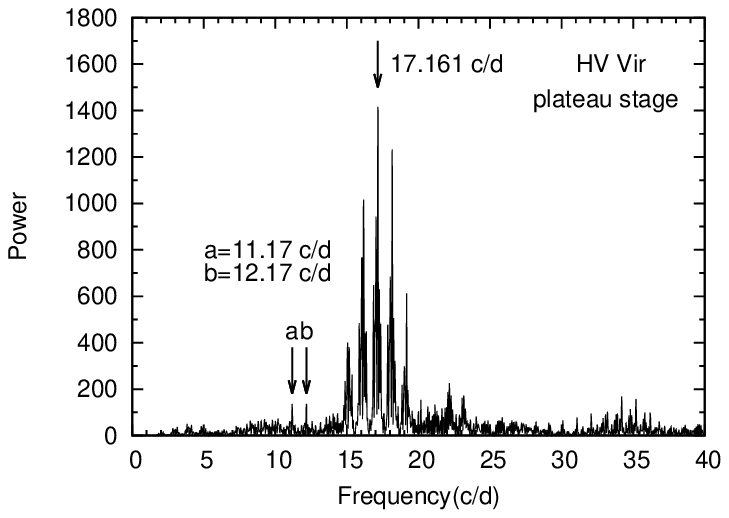}
\FigureFile(80mm,80mm){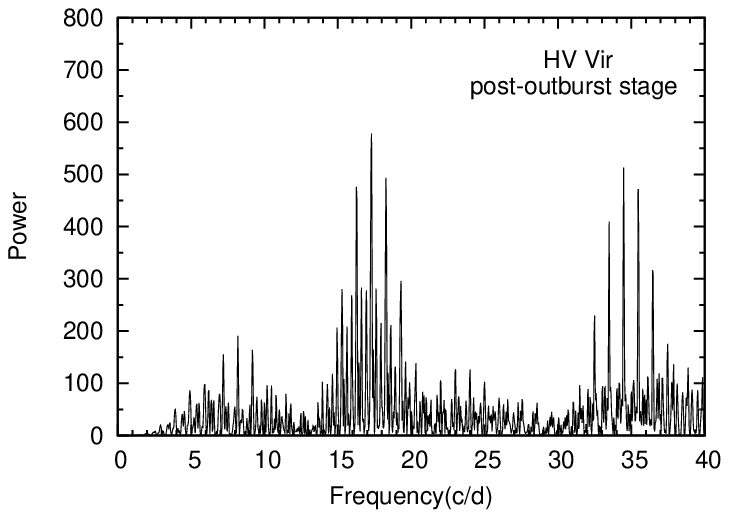}
\end{center}
\caption{(Left) Discrete Fourier Transform during the main plateau stage after
  the appearance of superhumps (pl.). The strongest signal corresponds to the
  mean superhump frequency of 17.16(4) c/d. Weak signals at 11.17(4) and 12.17(4) c/d are detectable. (Right) Discrete Fourier Transform after the end of the main superoutburst. No significant signals around ${\sim}$ 11 c/d was found.}
\label{hvvir_fou}
\end{figure}

Figure \ref{lc_hvvir} shows the light curves of HV Vir during our run. As can be seen in figure \ref{lc_hvvir}, the plateau stage lasted for 20 d, which was a usual value for WZ Sge-type dwarf novae \citep{kat15wzsge}. Although WZ Sge-type dwarf novae frequently show rebrightenings \citep{pat98egcnc}, the present observations provided no evidence for rebrightenings. The absence of rebrightenings is also confirmed in the 1992 and 2016 superoutbursts of the same object (\cite{lei94hvvir}; \cite{pdot9}). These superoutbursts can be categorized into the type-D classification (no rebrightenings) introduced by \citet{ima06tss0222} and \citet{kat15wzsge}.

Representative light curves in early phase of the superoutburst
are given in figure \ref{rep_hvvir}, in which the characteristic
features of early and ordinary superhumps are visible. As illustrated
in figure \ref{rep_hvvir}, superhumps appeared on 2008 February 20 (JD 2454517). This means that the superhumps appeared on the 6th day after the beginning of the superoutburst. In conjunction with the reported visual observations, the early superhumps of HV Vir lasted for 6 d. Superhumps sustained after the end of the plateau stage. This phenomenon is common among WZ Sge-type dwarf novae \citep{pdot}.

\subsection{period analyses of light curves}

We performed period analyses during the plateau stage. After removing the general trend of the light curve, we measured the maximum timings of early and ordinary superhumps. Tables \ref{bjdmaxhvviresh} and \ref{bjdmaxhvvirsh} show the timings of early and ordinary superhump maxima,respectively. A linear regression to the early superhump maxima yields

\begin{equation}
BJD(max) = 2454512.1503(12) + 0.057093(45) \times E1
\end{equation}

where $E1$ is the cycle count since the appearance of early superhumps in our data. As for stage A (JD 2454517.10$-$18.34), a linear regression to the superhump maxima yields

\begin{equation}
BJD(max) = 2454517.1488(6) + 0.058533(58) \times E2
\end{equation}

where $E2$ is the cycle count since the appearance of ordinary superhumps in our data. Similarly, linear regression to the superhump maxima for stage B (JD 2454519.25$-$26.21) and C (after JD 2454527.05) yield 

\begin{equation}
BJD(max) = 2454517.1188(61) + 0.058500(47) \times E2
\end{equation}

for stage B, and

\begin{equation}
BJD(max) = 2454517.2149(50) + 0.057905(20) \times E2
\end{equation}

for stage C, respectively. Based on these results, we derived the mean periods of early and ordinary superhumps to be 0.057093(45) d for early superhumps, 0.058533(59) d for stage A superhumps, 0.058500(47) d for stage B superhumps, and 0.057905(20) d for stage C superhumps, respectively. The obtained early superhump period is in agreement with that reported during the 1992 superoutburst ($P_{\rm Esh}$ = 0.057085(23) d, \citet{kat01hvvir}). Also, the mean superhump period during stage A is again in agreement with that during the 2002 superoutburst ($P_{\rm sh}$ = 0.05844(24) d, \citet{ish03hvvir}).
According to \citet{pdot}, fractional period excesses during stage A over the mean superhump period during stage B tend to be clustered around 1.0-1.5$\%$. 
In the present case, however, the mean superhump periods did not differ between stage A and B. This may be due to the fact that our observations of stage B were clustered to the late stage of the superoutburst plateau. As noted in introduction, the stage B superhump period increases as the superoutburst proceeds. Therefore, the mean superhump period during stage B that we obtained in this paper might be overestimated. We summarize the results in table \ref{hvvir_period}. 

As noted in introduction, quiescent light curves showed a frequency of
11.20 c/d or 12.21 c/d \citep{pat03suumas}. In order to check whether these signals are present during our observations, we performed Discrete Fourier Transform analyses. Figure \ref{hvvir_fou} shows the results of our analyses. During the plateau stage after the appearance of the superhumps (pl.), we detected weak signals at 11.17(4) and 12.17(5) c/d (0.08953 d and 0.08217 d), which are close to those obtained in \citet{pat03suumas}. Similar signals were detected in e.g., FS Aur \citep{tov03fsaur}, GW Lib \citep{wou02gwlib}, and V455 And \citep{ara05v455and}, and SSS J122221.7-311525 \citep{neu17j1222}. It should be noted that these periods are significantly longer than the orbital ones. On the other hand, we did not detect these signals after the end of the superoutburst, despite the fact that such signals were detected during quiescence in other systems. Because little is known about such a long period of WZ Sge-type dwarf novae, we avoid further discussion on these periodicities.\footnote{See also \citet{2007ApJ...655..466T} for interpretation of these periodicities.}

\begin{figure*}
\begin{center}
\FigureFile(50mm,60mm){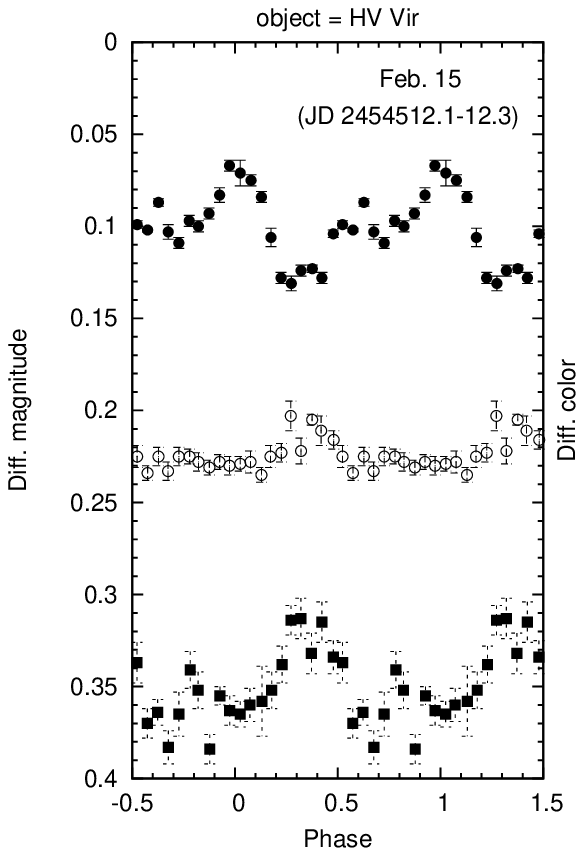}
\FigureFile(50mm,60mm){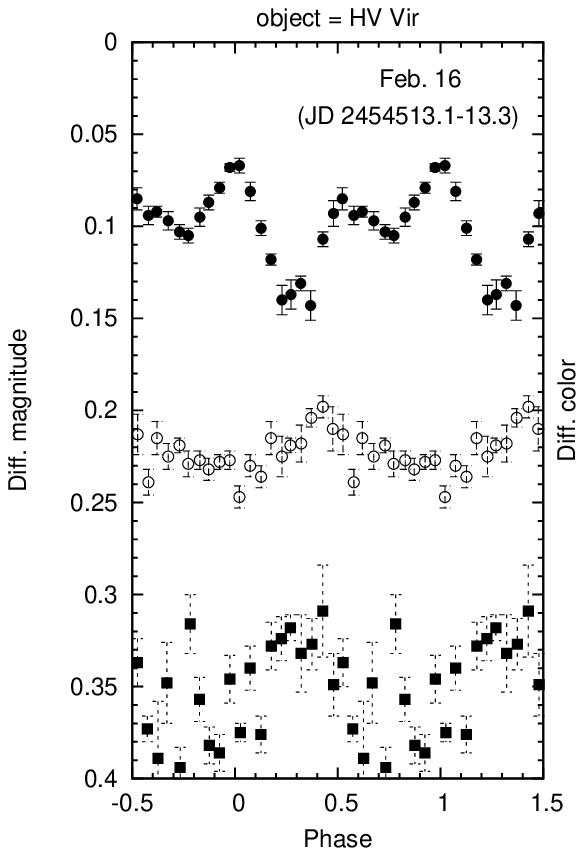}
\FigureFile(50mm,60mm){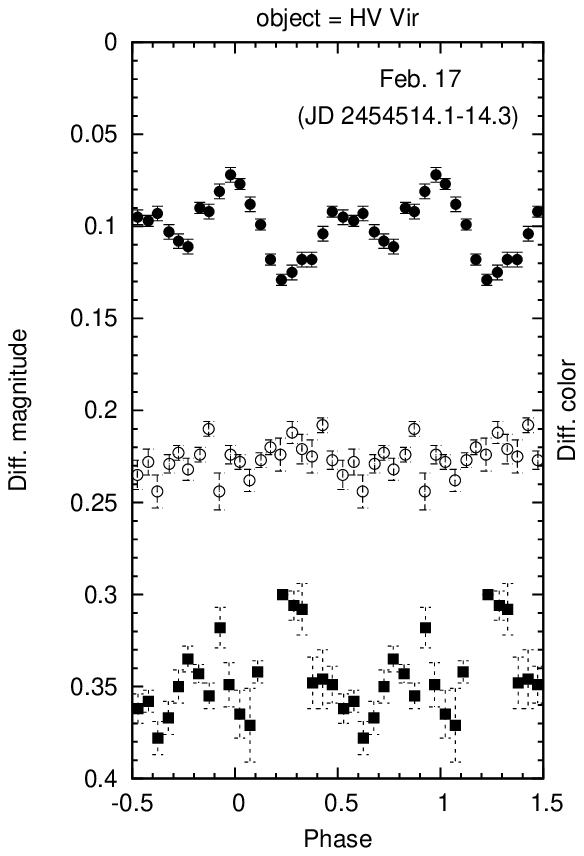}
\FigureFile(50mm,60mm){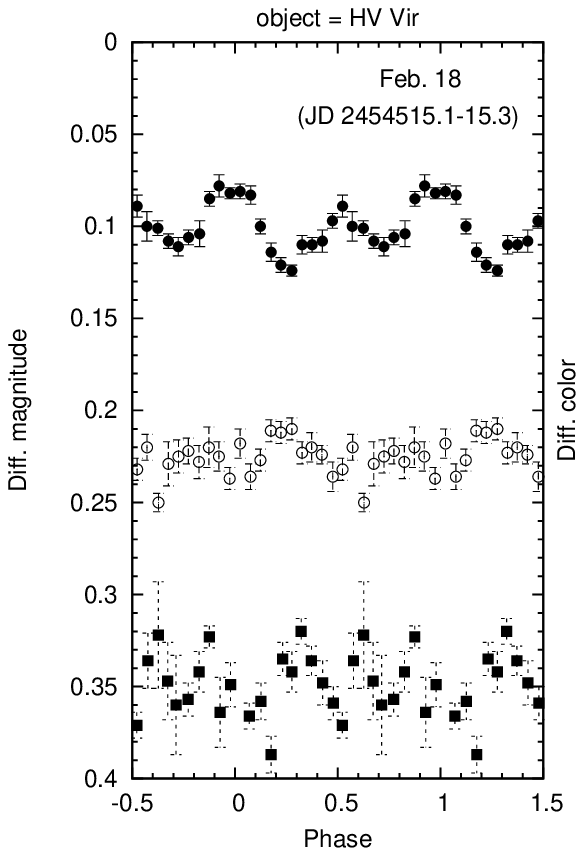}
\FigureFile(50mm,60mm){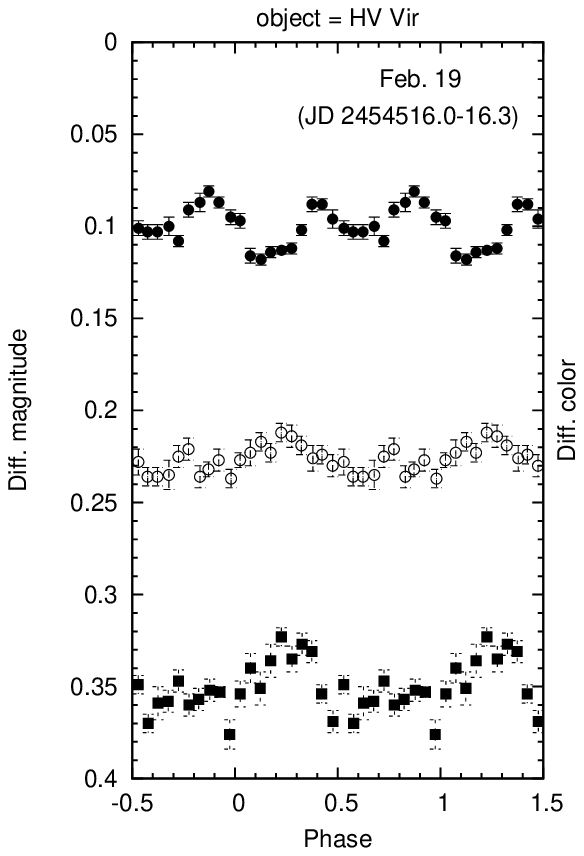}
\end{center}
\caption{Light curves and color variations of HV Vir during early superhump stage (Esh). Each panel shows $R_{\rm c}$ band (filled circles), $g'-I_{\rm c}$ (open circles), and $g'-K_{\rm s}$ (filled squares). These data are folded with 0.057093 d and the epoch is set on BJD 2454512.2100. The profiles of the light curves are double-peaked, characteristic of early superhumps. On the other hand, the color variations tend to show single-peaked profiles.}
\label{hvvir_Esh}
\end{figure*}

\subsection{profile of light curves and color variations}

Figure \ref{hvvir_Esh} shows $R_{\rm c}$ light curves and colors folded with $P_{\rm Esh}$ = 0.057093 d. Double-peaked modulations characteristic of early superhumps are conspicuous in each panel. In contrast to these light curves, color variations tend to show single-peaked profiles. It should be noted that colors are the bluest at the brightness minima. This property has been reported during the 2007 superoutburst of V455 And \citep{mat09v455and}.

\begin{figure}
\begin{center}
\FigureFile(40mm,60mm){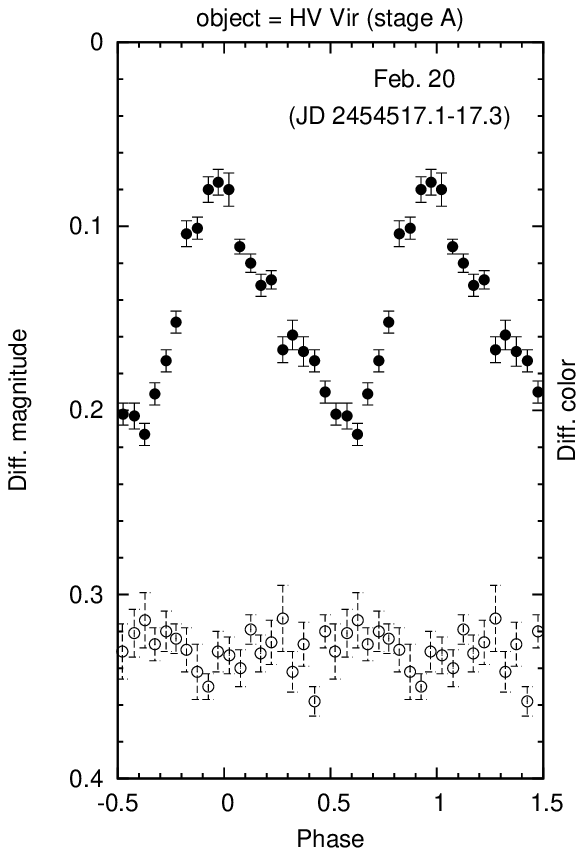}
\FigureFile(40mm,60mm){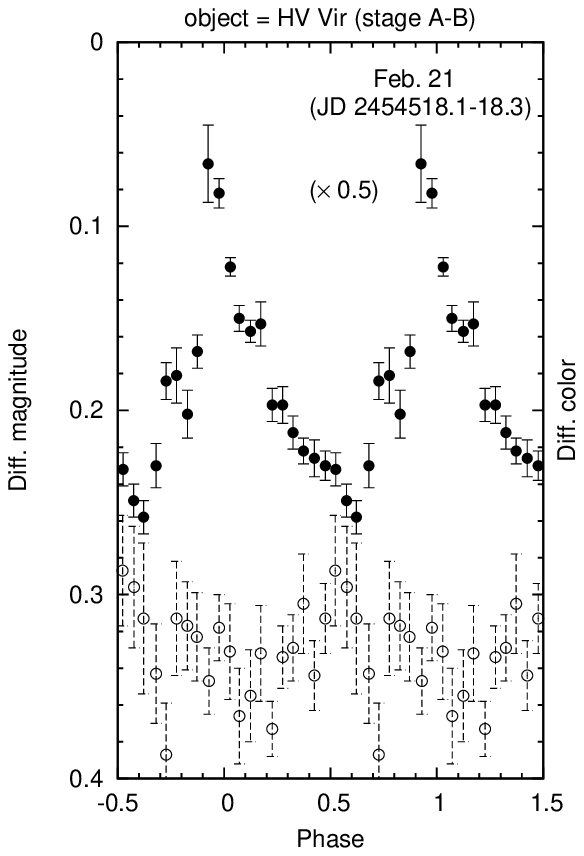}
\FigureFile(40mm,60mm){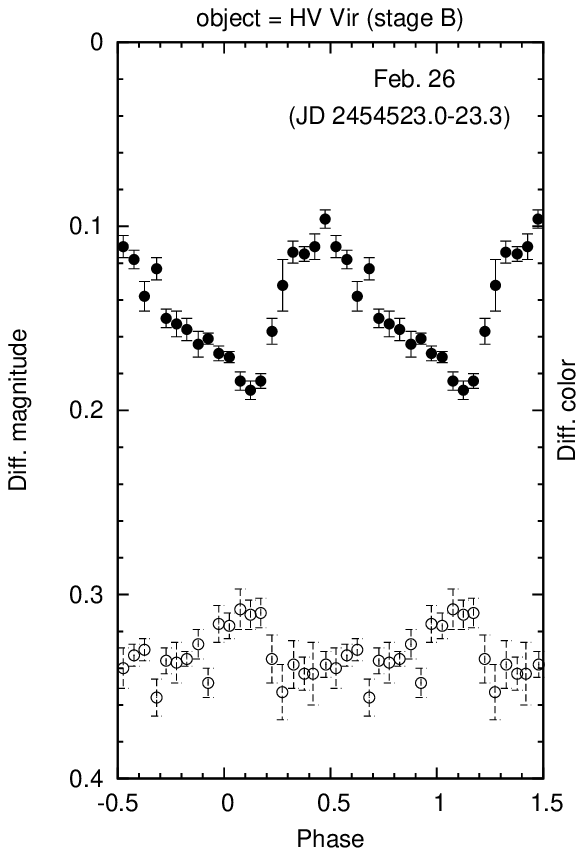}
\FigureFile(40mm,60mm){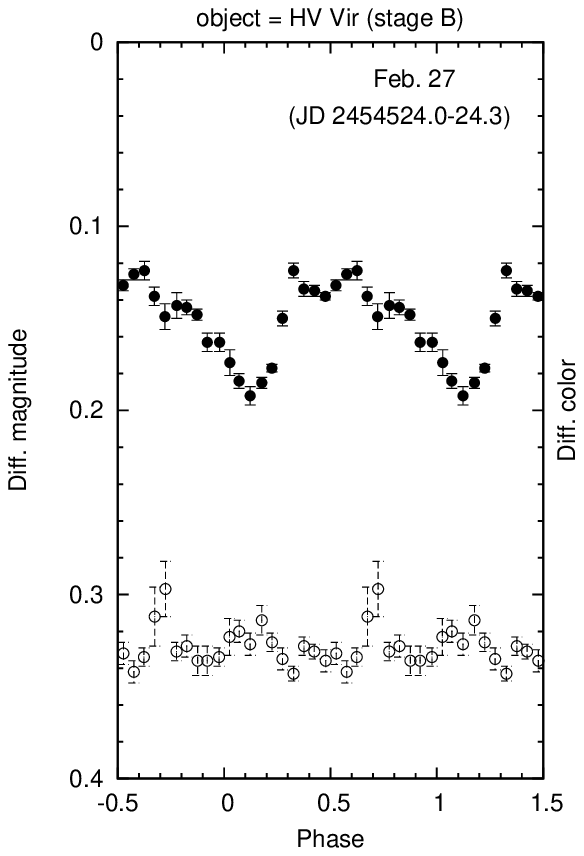}
\FigureFile(40mm,60mm){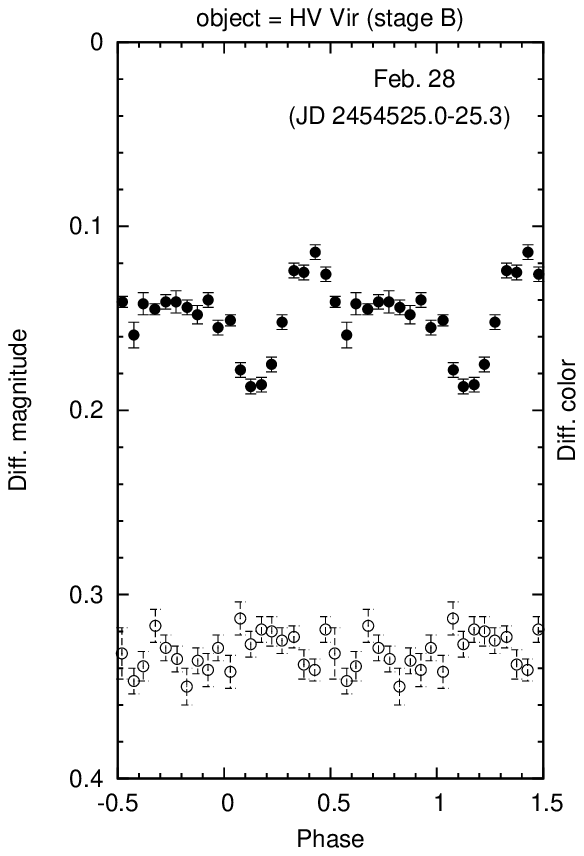}
\FigureFile(40mm,60mm){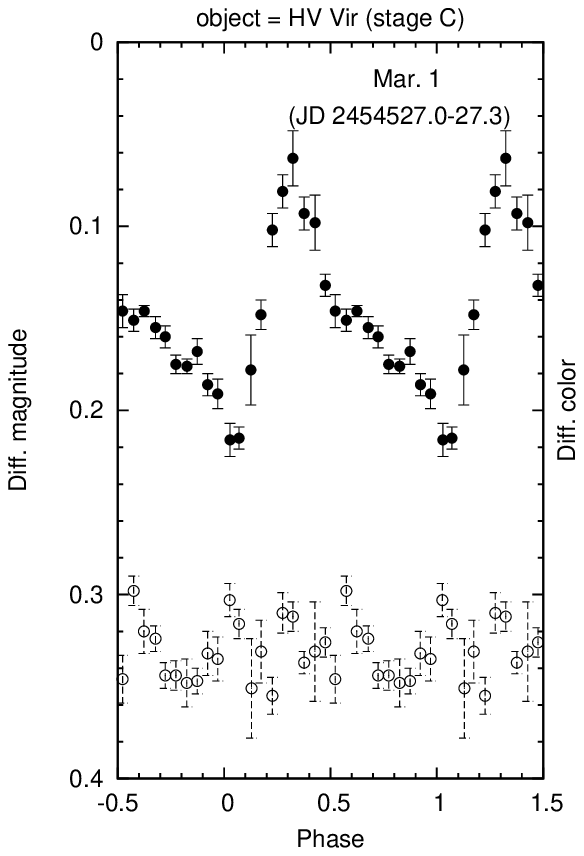}
\end{center}
\caption{$R_{\rm c}$ light curves (filled circles) and $g'-I_{\rm c}$ variations (open circles) of HV Vir during the plateau stage. We folded them with 0.058533 d on 2008 February 20 and 21, with 0.058500 d on February 26$-$28, and 0.057905 d on March 1, respectively. The epoch is set on BJD 2454517.1492. For display purpose, the superhump profile on February 21 is scaled down by 0.5.}
\label{hvvir_sh}
\end{figure}

Figure \ref{hvvir_sh} represents $R_{\rm c}$ light curves and $g'-I_{\rm c}$ color
variations folded with $P_{\rm sh}$ = 0.058533 d for February 20 and 21 (stage A), $P_{\rm sh}$ = 0.058500 d for February 26$-$28 (stage B), and $P_{\rm sh}$ = 0.058394 d for March 1 (stage C). We can see a tendency that $g'-I_{\rm c}$ colors are the bluest at the brightness minima on February 21 and 26. As was observed in HV Vir, the difference in phase between magnitudes and colors was also observed in other superhumps, such as e.g., VW Hyi \citep{sch80vwhyi}, EK TrA \citep{has85ektra}, and V455 And \citep{mat09v455and}.

\begin{table}
\caption{Timings of early superhump maxima of HV Vir.}
\begin{center}
\begin{tabular}{ccc}
\hline\hline
E1 & max & Error \\
\hline
0 & 4512.1526 & 0.0009 \\
1 & 4512.2092 & 0.0014 \\
2 & 4512.2602 & 0.0013 \\
3 & 4512.3206 & 0.0025 \\
17 & 4513.1236 & 0.0008 \\
18 & 4513.1776 & 0.0009 \\
35 & 4514.1450 & 0.0008 \\
36 & 4514.2076 & 0.0011 \\
37 & 4514.2630 & 0.0010 \\
53 & 4515.1766 & 0.0012 \\
\hline
\multicolumn{3}{l}{$^{\dagger}$BJD$-$2450000.} \\
\end{tabular}
\end{center}
\label{bjdmaxhvviresh}
\end{table}

\begin{table}
\caption{Timings of ordinary superhump maxima of HV Vir.}
\begin{center}
\begin{tabular}{cccc}
\hline\hline
E2 & max & Error & Stage \\
\hline
0 & 4517.1496 & 0.0007 & A \\
1 & 4517.2059 & 0.0005 & A \\
2 & 4517.2652 & 0.0005 & A \\
3 & 4517.3256 & 0.0006 & A \\
17 & 4518.1437 & 0.0008 & A-B \\
18 & 4518.2024 & 0.0014 & A-B \\
103 & 4523.1414 & 0.0008 & B \\
105 & 4523.2608 & 0.0009 & B \\
120 & 4524.1415 & 0.0012 & B \\
122 & 4524.2545 & 0.0009 & B \\
123 & 4524.3167 & 0.0012 & B \\
137 & 4525.1338 & 0.0015 & B \\
138 & 4525.1949 & 0.0014 & B \\
139 & 4525.2490 & 0.0010 & B \\
140 & 4525.3089 & 0.0008 & B \\
155 & 4526.1833 & 0.0014 & B \\
171 & 4527.1171 & 0.0006 & C \\
174 & 4527.2892 & 0.0012 & C \\
175 & 4527.3481 & 0.0010 & C \\
225 & 4530.2456 & 0.0007 & C \\
226 & 4530.3033 & 0.0007 & C \\
240 & 4531.1088 & 0.0042 & C \\
241 & 4531.1722 & 0.0017 & C \\
242 & 4531.2293 & 0.0016 & C \\
243 & 4531.2890 & 0.0009 & C \\
261 & 4532.3216 & 0.0030 & C \\
275 & 4533.1356 & 0.0027 & C \\
277 & 4533.2561 & 0.0015 & C \\
278 & 4533.3081 & 0.0023 & C \\
294 & 4534.2415 & 0.0017 & C \\
295 & 4534.3006 & 0.0020 & C \\
\hline
\multicolumn{4}{l}{$^{\dagger}$BJD$-$2450000.} \\
\end{tabular}
\end{center}
\label{bjdmaxhvvirsh}
\end{table}

\begin{table}
\caption{Results of period analyses of HV Vir.}
\begin{center}
\begin{tabular}{ccc}
\hline\hline
Stage$^*$ & JD(start-end)\commentb & period(day) \\
\hline
Esh & 4511.2348$-$4516.3685 & 0.057093(45) \\
A &   4517.1018$-$4518.3470 & 0.058533(58) \\
B &   4518.1024$-$4526.2114 & 0.058500(47) \\
C &   4527.0560$-$4541.2899 & 0.057905(20) \\
\hline
\multicolumn{3}{l}{$^*$ Esh: Early superhumps.} \\
\multicolumn{3}{l}{A: stage A. B: stage B. C: stage C} \\
\multicolumn{3}{l}{\commentb JD$-$2450000.} \\
\end{tabular}
\end{center}
\label{hvvir_period}
\end{table}

\subsection{amplitudes of early and ordinary superhumps}

\begin{figure}
\begin{center}
\FigureFile(80mm,80mm){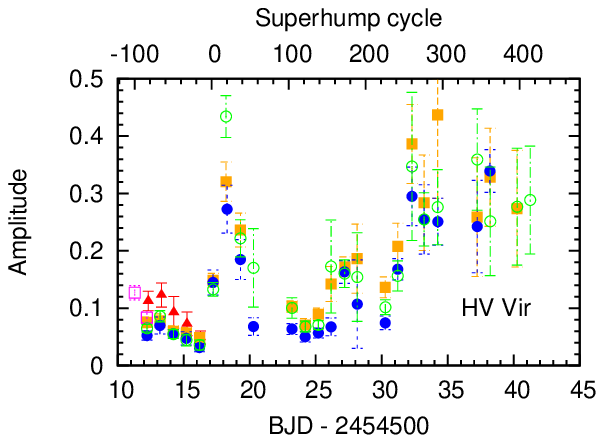}
\FigureFile(80mm,80mm){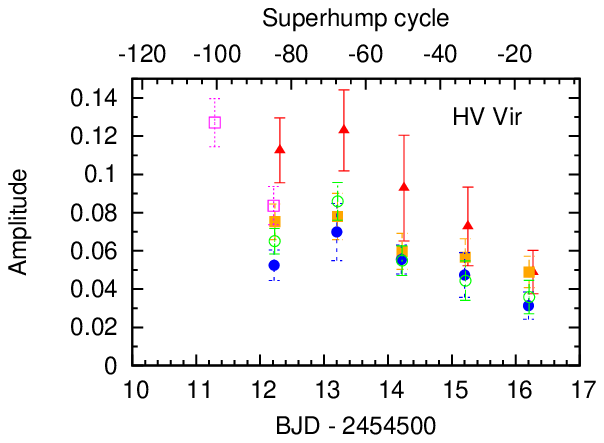}
\end{center}
\caption{Left: Amplitudes of early and ordinary superhumps of HV Vir. Filled circles (blue), open circles (green), and filled squares (orange) represent data of $g'$, $R_{\rm c}$, and $I_{\rm c}$ band, respectively. Near-infrared data are marked with open squares ($H$ band, magenta) and filled triangles ($K_{\rm s}$ band, red), respectively. The amplitudes of superhumps reach a maximum at around the stage A$-$B transition, after which the amplitudes are decayed until around the stage B$-$C transition (around JD 2454526). Right: Enlarged figure during the early superhump stage. Note that near-infrared amplitudes are larger than optical amplitudes.}
\label{hvvir_amp}
\end{figure}

We studied nightly-averaged amplitudes of each band, which is
illustrated in figure \ref{hvvir_amp}. It should be noted that the mean
amplitudes of early superhumps in near-infrared bands are larger than those in the optical ranges. We can see a tendency that the amplitudes of $g'$ band show small values over the course of the superoutburst. The amplitudes of early superhumps decreased as the superoutburst proceeded. After the appearance of the superhumps, the maximum of the amplitude occurred on JD 2454518, corresponding to the transition between the stage A and B. In contrast with the amplitudes of early superhumps, no significant dependency on wavelength was found in ordinary superhumps.

\section{Results of OT J012059.6+325545}

\subsection{overall light curves}

\begin{figure}
\begin{center}
\FigureFile(80mm,80mm){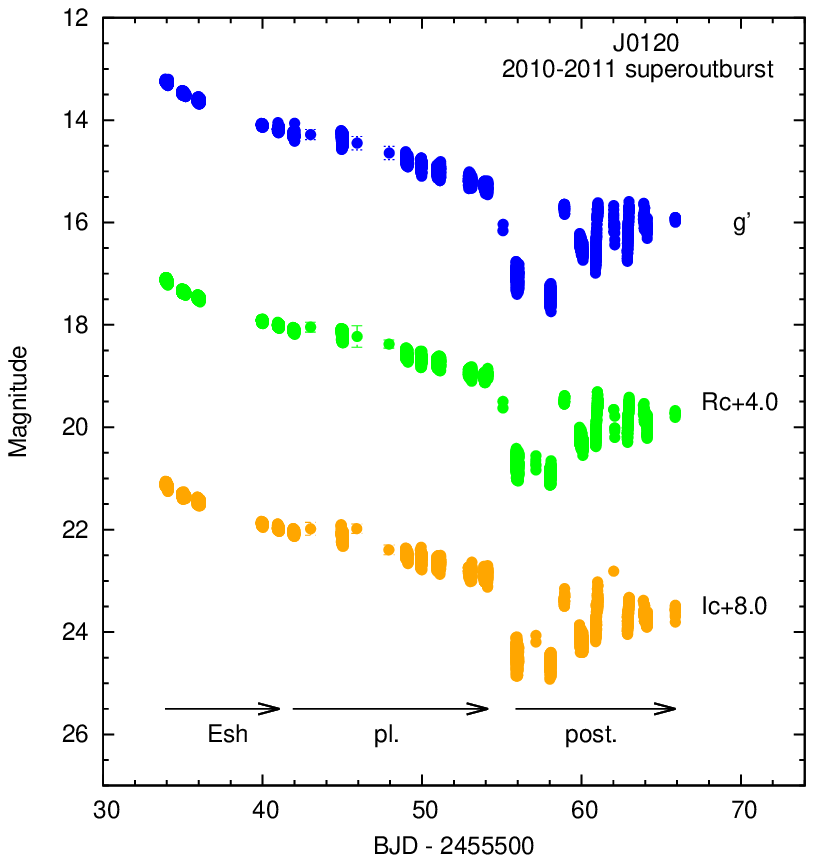}
\end{center}
\caption{$g'$ (blue), $R_{\rm c}$ (green) and $I_{\rm c}$ (orange) light curves of J0120 during the 2010-2011 superoutburst. In combination with the discovery of the superoutburst, the duration of the plateau stage is estimated as long as 25 d. Each light curve clearly shows a "dip'' after the end of the plateau stage.}
\label{lc_j0120}
\end{figure}

\begin{figure}
\begin{center}
\FigureFile(80mm,160mm){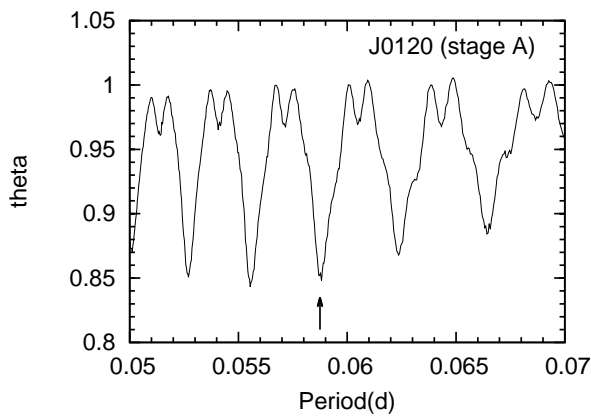}
\end{center}
\caption{Theta diagram between JD 2455540.9302$-$43.1383, corresponding to the onset of the growing superhumps. Considering the mean early superhump period of J0120 ($P_{\rm Esh}$ = 0.057147 d), the best candidate period is 0.05888 d.}
\label{j0120_pdmA}
\end{figure}

\begin{figure}
\begin{center}
\FigureFile(80mm,80mm){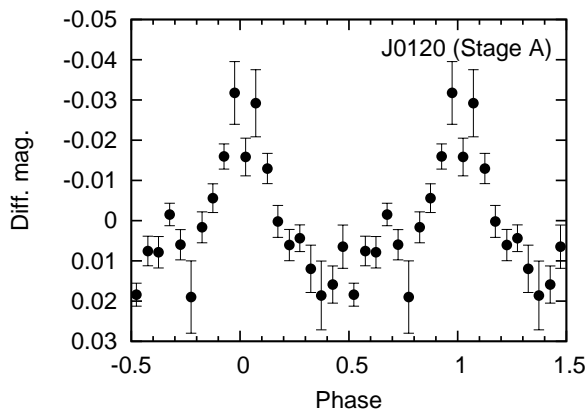}
\end{center}
\caption{Phase-averaged light curve on 2010 December 10$-$12 folded with 0.05875 d. The single-peaked light curve is indicative of the growing superhumps.}
\label{stageA_j0120}
\end{figure}

\begin{table}
\caption{Timings of early superhump maxima of J0120.}
\begin{center}
\begin{tabular}{ccc}
\hline\hline
E1 & max & Error \\
\hline
0 & 5533.9262 & 0.0007 \\
1 & 5533.9831 & 0.0006 \\
2 & 5534.0389 & 0.0013 \\
18 & 5534.9529 & 0.0007 \\
19 & 5535.0144 & 0.0014 \\
35 & 5535.9255 & 0.0009 \\
36 & 5535.9803 & 0.0010 \\
37 & 5536.0407 & 0.0009 \\
105 & 5539.9244 & 0.0023 \\
106 & 5539.9852 & 0.0017 \\
\hline
\multicolumn{3}{l}{$^{\dagger}$BJD$-$2450000.} \\
\end{tabular}
\end{center}
\label{bjdmaxj0120esh}
\end{table}

\begin{table}
\caption{Timings of ordinary superhump maxima of J0120.}
\begin{center}
\begin{tabular}{cccc}
\hline\hline
E2 & max & Error & Stage \\
\hline
0 & 5540.9534 & 0.0010 & A \\
17 & 5541.9533 & 0.0011 & A \\
18 & 5542.0099 & 0.0015 & A \\
68 & 5544.9258 & 0.0007 & B \\ 
69 & 5544.9854 & 0.0004 & B \\ 
70 & 5545.0418 & 0.0004 & B \\
139 & 5549.0247 & 0.0007 & B \\ 
155 & 5549.9509 & 0.0009 & B \\ 
172 & 5550.9305 & 0.0006 & B \\ 
173 & 5550.9888 & 0.0011 & B \\
174 & 5551.0452 & 0.0008 & B \\
175 & 5551.1026 & 0.0019 & B \\
\hline
\multicolumn{4}{l}{$^{\dagger}$BJD$-$2450000.} \\
\end{tabular}
\end{center}
\label{bjdmaxj0120sh}
\end{table}

\begin{table}
\caption{Results of period analyses of J0120.}
\begin{center}
\begin{tabular}{ccc}
\hline\hline
Stage$^*$ & JD(start-end)\commentb & period(day) \\
\hline
Esh & 5533.8857$-$5541.0535 & 0.057147(15) \\
A &   5540.9302$-$5543.1383 & 0.05875(11) \\
B &   5544.9190$-$5554.1251 & 0.057729(8) \\
\hline
\multicolumn{3}{l}{$^*$ Esh: Early superhumps.} \\
\multicolumn{3}{l}{A: stage A. B: stage B.} \\
\multicolumn{3}{l}{\commentb JD$-$2450000.} \\
\end{tabular}
\end{center}
\label{j0120_pdmT}
\end{table}

Figure \ref{lc_j0120} shows $g'$, $R_{\rm c}$, and $I_{\rm c}$ band
light curves. The light curves closely resemble that observed during the 2001 superoutburst of WZ Sge, in terms of the amplitude of the superoutburst (${\sim}$ 8 mag), the duration of the plateau stage (${\sim}$ 25 d), and the rebrightenings. The rebrightenings are classified as the type A-B rebrightenings defined by \citet{ima06tss0222} and \citet{kat15wzsge}. We detected rising stages of the second and third rebrightenings on JD 2455561 and JD 2455563, respectively. The duration of each rebrightening was about 2 d, which is almost the same as that observed during the 2001 superoutburst of WZ Sge itself \citep{pdot}. We derived the rising rates of the second and third rebrightenings in $R_{\rm c}$ band to be 5.74(11) mag/d and 5.56(14) mag/d, respectively. These values indicate the outside-in type of these rebrightenings \citep{kat04egcnc}.

\subsection{period analyses of light curves}

In order to determine the mean periods of early and ordinary superhumps, we calculated maximum timings of humps, which are tabulated in table \ref{bjdmaxj0120esh} and \ref{bjdmaxj0120sh}. A linear regression to the early superhump maxima yields

\begin{equation}
BJD(max) = 2455533.9256(8) + 0.057147(15) \times E1,
\end{equation}

where $E1$ is the cycle count since the appearance of early superhumps in our data. Based on the regression analysis, we derived the mean early superhump period to be $P_{\rm Esh}$ = 0.057147(15) d.

\citet{nak13j0120} reported that a hint of growing superhumps (stage A superhumps) was observed around JD 2454442.0. In order to confirm this, we performed the phase dispersion minimization method (PDM, \citet{pdm}) on JD 2455540.93$-$43.13. The resulting theta diagram is shown in figure \ref{j0120_pdmA}. In this diagram, we can detect several peaks, of which $P$ = 0.05888(11) d is the best candidate period, based on the fact that the mean superhump period is slightly longer than the orbital period (or $P_{\rm Esh}$) of the system \citep{kat15wzsge}. Our regression analyses yield

\begin{equation}
BJD(max) = 2455540.9535(15) + 0.05875(11) \times E2
\end{equation}

for JD 2455540.93$-$43.13, and

\begin{equation}
BJD(max) = 2455540.9434(11) + 0.057729(8) \times E2
\end{equation}

for JD 2455544.91$-$54.12, respectively. The estimated period between 
JD 2455540.93$-$43.13 ($P_{\rm sh}$ = 0.05875(11) d) is in agreement with that derived by the PDM ($P_{\rm sh}$ = 0.05888(11) d). On the other hand, the mean superhump period for JD 2455544.91$-$54.12 was shorter than that obtained by \citet{nak13j0120} ($P_{\rm sh}$ = 0.057814(12) d) by 0.15 \%. This difference may be caused by the fact that the different baselines were used for determing the mean superhump period, and the fact that $P_{\rm sh}$ varies as the superoutburst proceeds \citep{pdot}. It should be noted that the mean superhump period between JD 2455540.93$-$43.13 is about 1.8\% longer than that between JD 2455544.91$-$54.12. Such a long periodicity after the end of early superhumps is characteristic of stage A superhumps \citep{pdot}. Figure \ref{stageA_j0120} shows $R_{\rm c}$ light curve folded with $P_{\rm sh}$ = 0.05875 d, in which a single-peaked modulation, characteristic of superhumps, is visible. Therefore, it is likely that the observed modulations between JD 2455540.93$-$43.13 are indeed stage A superhumps. We summarize our period analyses of J0120 in table \ref{j0120_pdmT}.

\begin{figure*}
\begin{center}
\FigureFile(40mm,60mm){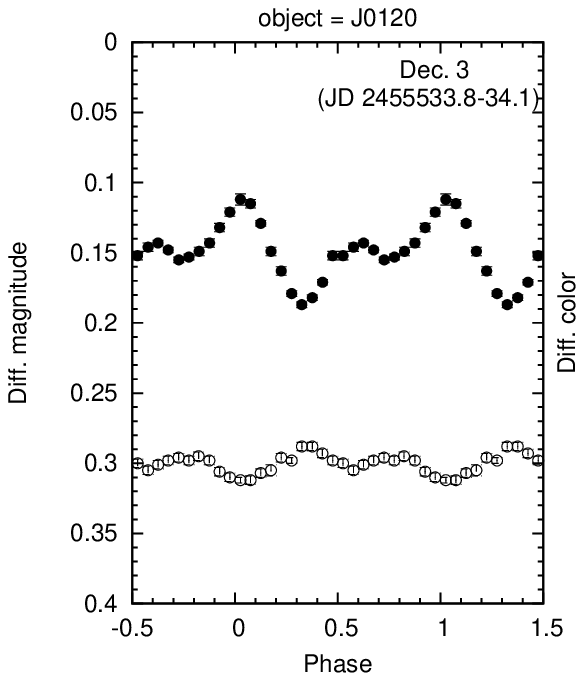}
\FigureFile(40mm,60mm){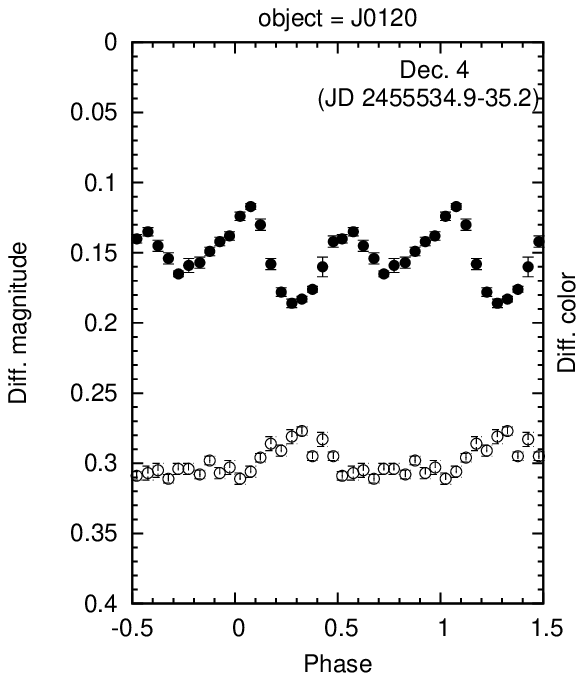}
\FigureFile(40mm,60mm){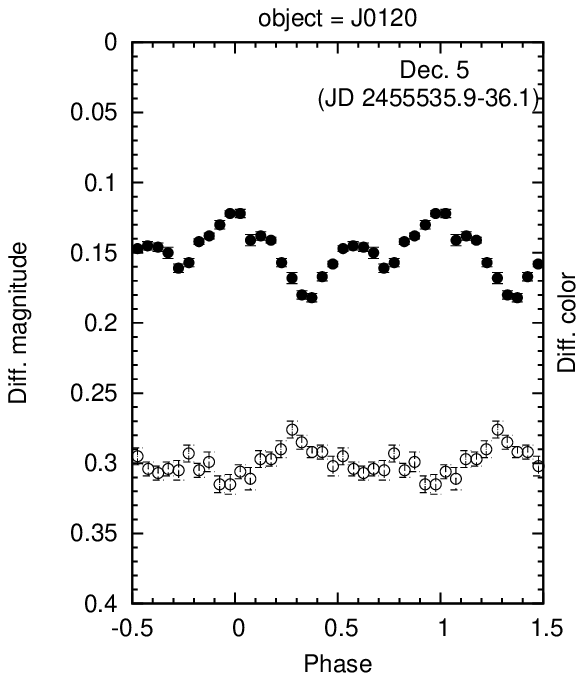}
\FigureFile(40mm,60mm){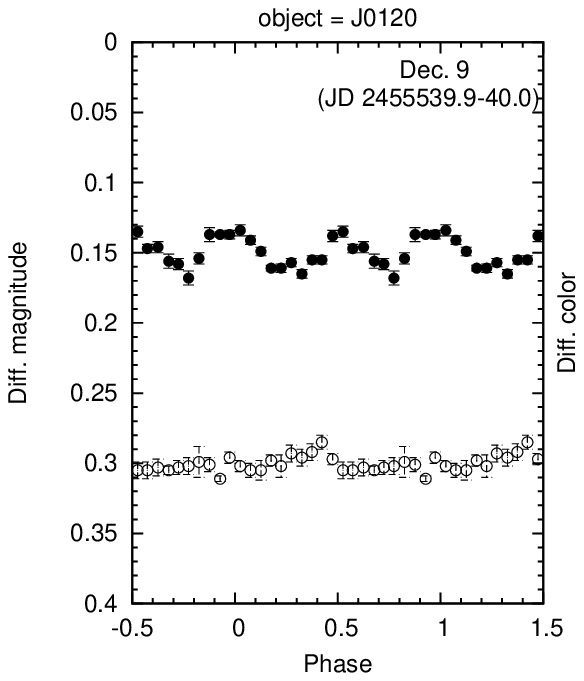}
\FigureFile(40mm,60mm){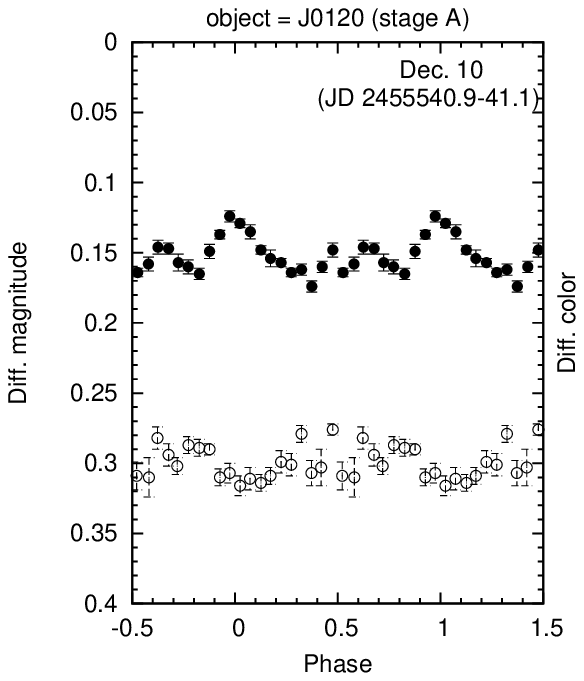}
\FigureFile(40mm,60mm){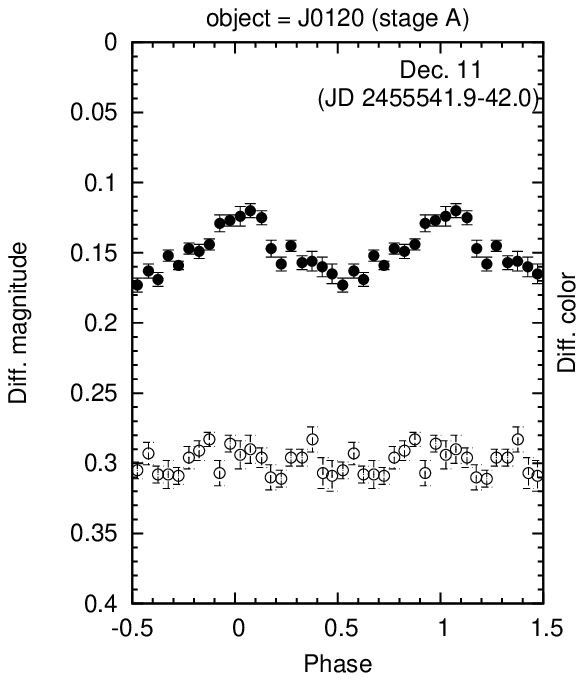}
\FigureFile(40mm,60mm){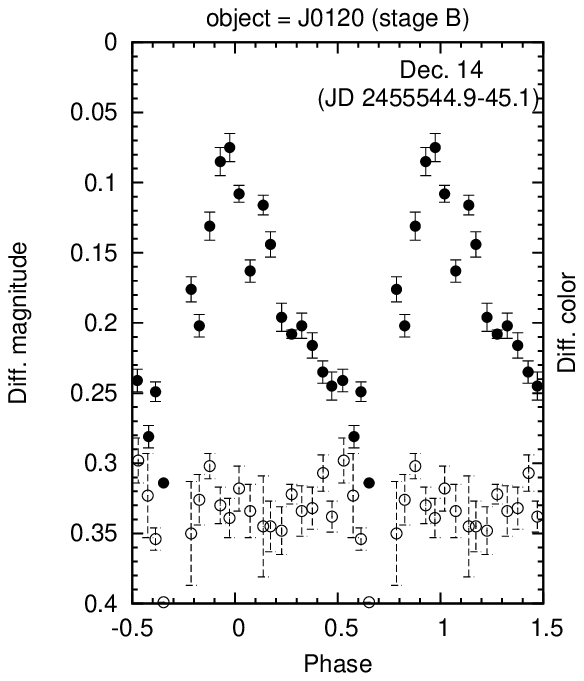}
\FigureFile(40mm,60mm){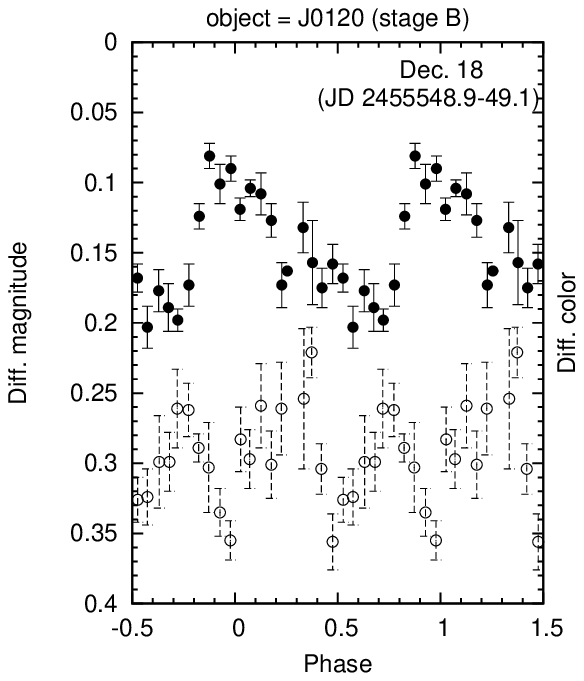}
\FigureFile(40mm,60mm){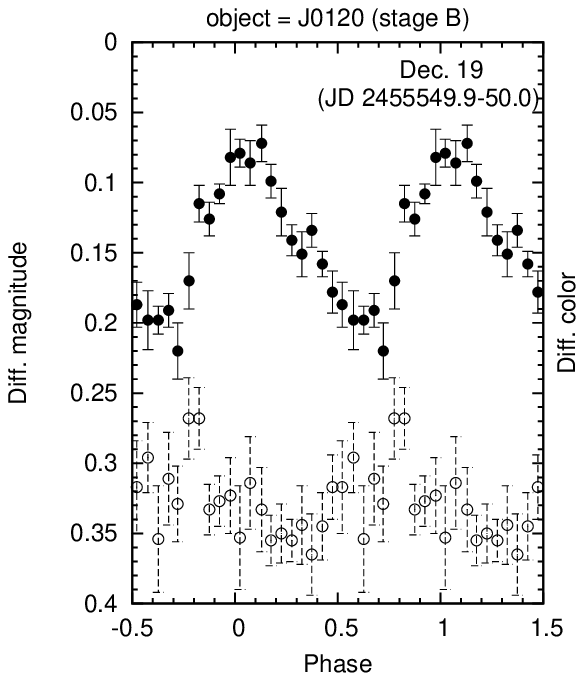}
\FigureFile(40mm,60mm){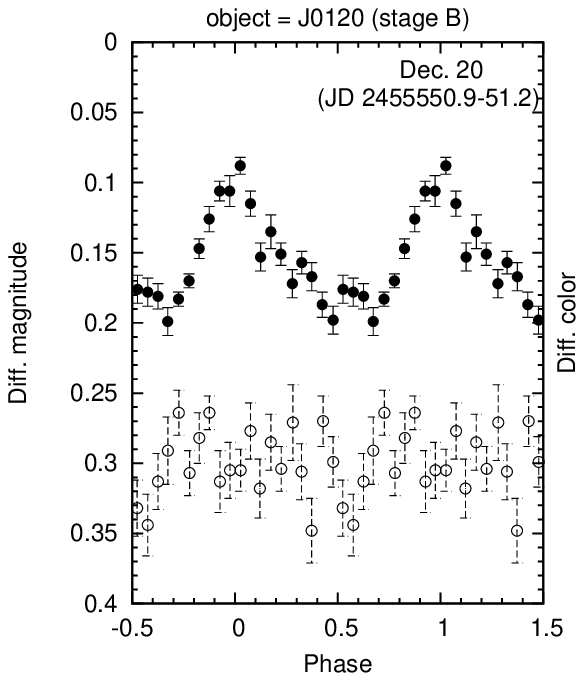}
\FigureFile(40mm,60mm){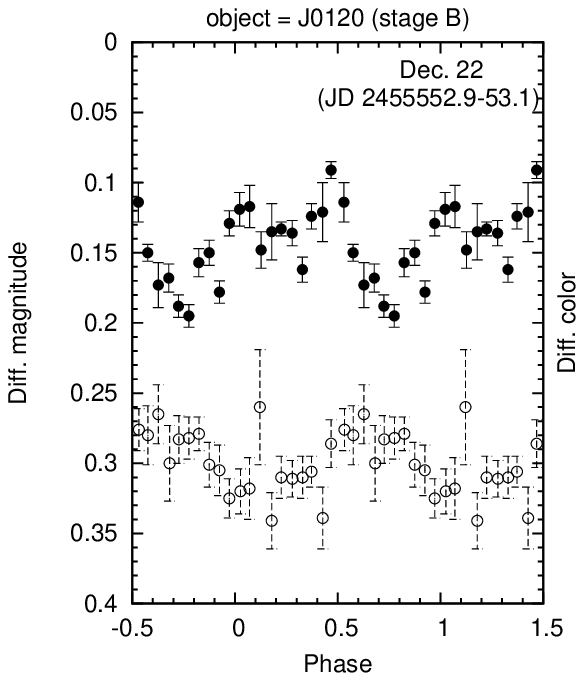}
\FigureFile(40mm,60mm){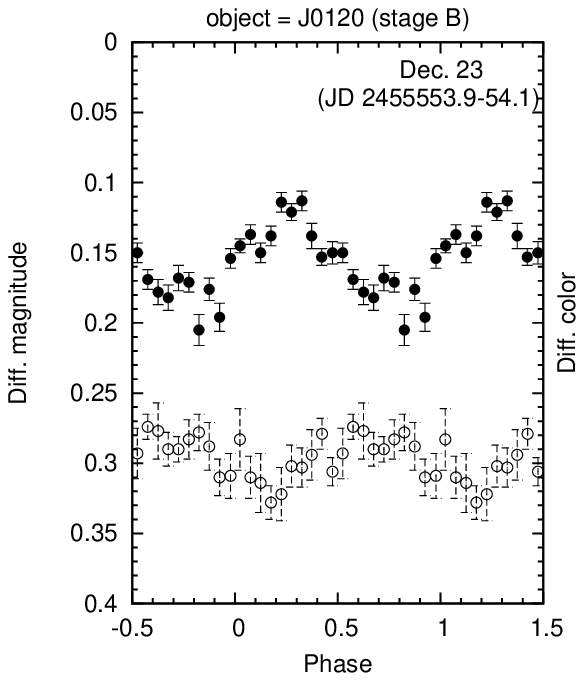}
\end{center}
\caption{$R_{\rm c}$ light curves (filled circles)
  and $g'-I_{\rm c}$ variations (open circles). These data are folded
  with 0.057147 d (December 3$-$9), 0.05875 d (December 10$-$11) and 0.057834 d (after December 14),
  respectively. The epoch is set on BJD 2455544.4060.}
\label{j0120_shvar}
\end{figure*}

\citet{kat13qfromstageA} developed a new method for estimating the mass ratio of the system using the orbital and stage A superhump periods. Empirically, $P_{\rm Esh}$ is identical with the orbital period of the system \citep{kat15wzsge}. Here we regard $P_{\rm Esh}$ as $P_{\rm orb}$ and estimated the mass ratio to be $q$ = 0.073(7). The error of the mass ratio is estimated based on the analytic formulae of \citet{kat13qfromstageA}, the error of the stage A superhump period, and that of the early superhump period. The estimated mass ratio is common among WZ Sge-type dwarf novae \citep{kat15wzsge}. The validity of the obtained mass ratio should be tested by future spectorscopic observations.

\subsection{profiles of humps and color variations}

Figure \ref{j0120_shvar} shows $R_{\rm c}$ nightly-averaged light
curves and $g'-I_{\rm c}$ variations folded with the above obtained periods. Here we precluded several nights data with low quality and short observations. Although the profiles of early superhumps vary from night to night, double-peaked light curves are visible until 2010 December 9 (JD 2455540.02). In the most of the panels during the early superhump stage, the bluest peak in $g'-I_{\rm c}$ occurs when $R_{\rm c}$ is the faintest, similar to that observed in HV Vir and other WZ Sge-type dwarf novae (\cite{mat09v455and}; \cite{neu17j1222}). The $g'-I_{\rm c}$ color variations of ordinary superhumps are not noticeable compared with those of early superhumps, but the bluest peaks correspond to the superhump minima in 2010 December 19 (JD 2455549.89$-$49.99) 22 (JD 2455552.89$-$53.12), and 23 (JD 2455553.87$-$54.12).

\begin{figure}
\begin{center}
\FigureFile(80mm,80mm){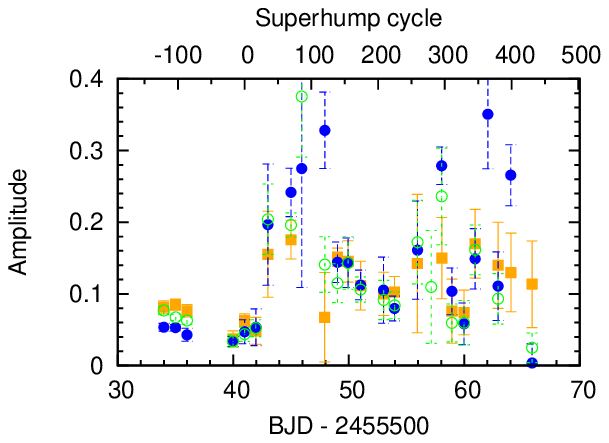}
\FigureFile(80mm,80mm){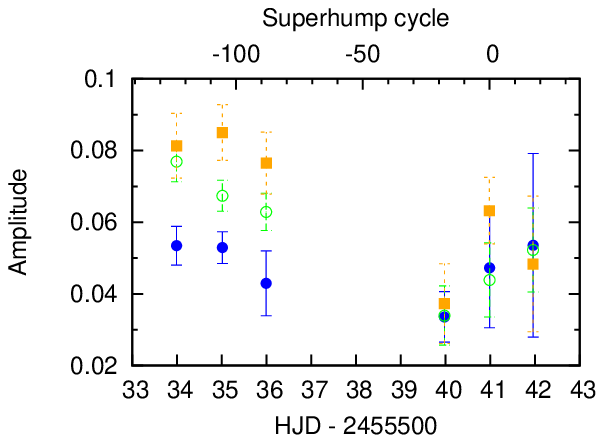}
\end{center}
\caption{Left: Amplitudes of early and ordinary superhumps of J0120. Filled circles (blue), open circles (green), and filled squares (orange) represent data of $g'$, $R_{\rm c}$, and $I_{\rm c}$ band, respectively.
The maximum amplitude occurred around the stage A$-$B transition (around BJD 2455543), after which the amplitudes gradually decreased until the end of the plateau stage. Right: enlarged figure during the early superhump phase. The amplitudes of early superhumps decrease as the superoutburst proceeds. The amplitudes of early superhumps depend on wavelength, particularily on the first three nights of our observations.}
\label{j0120_amp}
\end{figure}

Figure \ref{j0120_amp} shows the evolution of amplitudes of variability in each band. It should be noted that, on the first three nights of our observations, the amplitudes of early superhumps are the largest in $I_{\rm c}$ band and the smallest in $g'$ band. The amplitudes of early superhumps decreased as the outburst proceeded. On the other hand, the amplitudes of ordinary superhumps seems to be independent of wavelength.

\section{Discussion}

\begin{figure*}
\begin{center}
\FigureFile(80mm,80mm){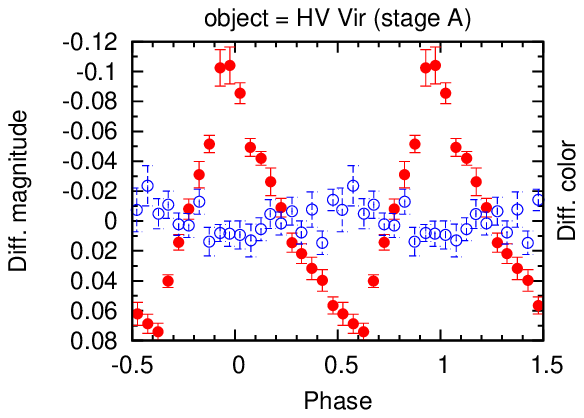}
\FigureFile(80mm,80mm){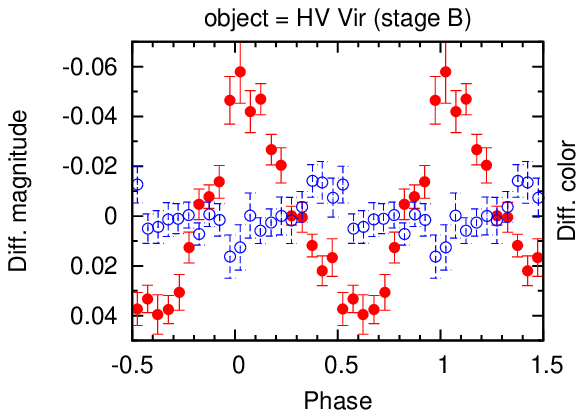}
\FigureFile(80mm,80mm){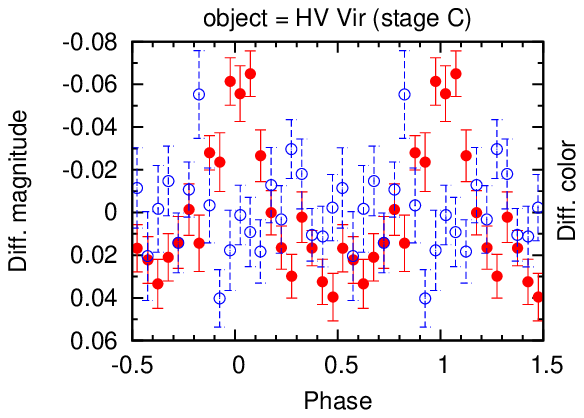}
\FigureFile(80mm,80mm){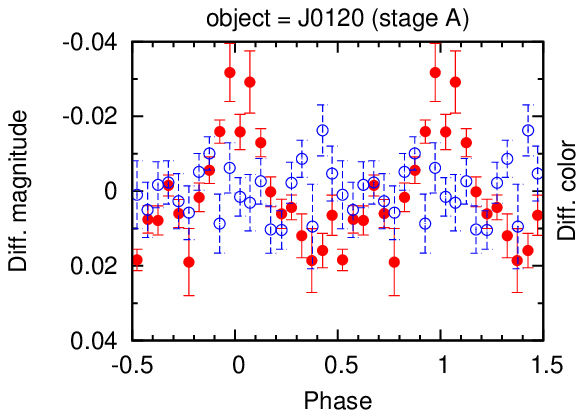}
\FigureFile(80mm,80mm){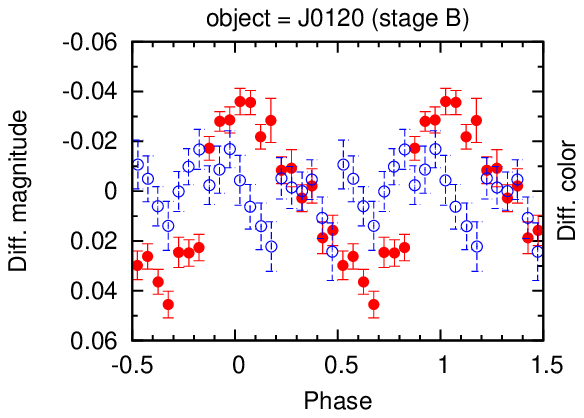}
\end{center}
\caption{Averaged profiles of superhumps and colors at different stages. Filled red circles represent superhumps in $R_{\rm c}$ band, while opened blue circles represent $g' - I_{\rm c}$ color variations.}
\label{shvarAB}
\end{figure*}

\begin{figure}
\begin{center}
\FigureFile(80mm,80mm){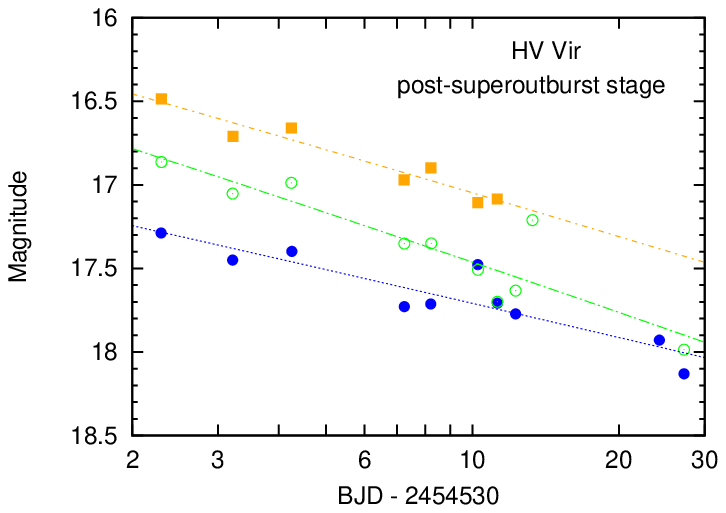}
\end{center}
\caption{$g'$ (blue), $R_{\rm c}$ (green), and $I_{\rm c}$ (orange) light curves of the post-superoutburst stage of HV Vir. The abscissa denote the date in logarithmic scale since BJD 2454530, corresponding to the beginning of the rapid declining stage, and the ordinate denotes the magnitude of each band. The linear lines in the figure mean the best-fitting power-law functions. Note that the $g'$ and $R_{\rm c}$ band light curves show moderate declines.}
\label{post_gri}
\end{figure}


As described in the previous section, the amplitudes of early superhumps are different in different bands. For example, $H$ and $K_{\rm s}$ amplitudes in HV Vir showed larger values compared with optical amplitudes, particularily in the first half of the early superhump stage. As for J0120, we can see a tendency that the amplitudes of early superhumps are the largest in $I_{\rm c}$ band and the smallest in $g'$ band for the first three nights of our observations. In general, $g'$ band amplitudes tend to show small values.

\begin{table}
\caption{$\Delta$ of each object at different stages.}
\begin{center}
\begin{tabular}{cccc}
\hline\hline
Object & stage A & stage B & stage C\\
\hline
HV Vir & 0.2 & 0.4 & 1 \\
J0120 & 0.6 & 0.4 & - \\
J1222 & 0.3 & 0.6${\sim}$1 & 1\\
\hline
\multicolumn{4}{l}{J1222 = SSS J122221.7$-$311525} \\
\end{tabular}
\end{center}
\label{delta}
\end{table}

Although the working mechanism on early superhumps is still in debate, theoretical works suggest that the outer region of the accretion disk plays a significant role in producing early superhumps (\cite{nog97alcom}; \cite{kat02wzsgeESH}; \cite{uem12ESHrecon}). From the observational point of view, the light sources of early superhumps are likely to have dependence on the system inclination. For example, only very low-amplitude modulations were seen at the early phase of the 2007 superoutburst of GW Lib \citep{pdot}, whose inclination is estimated as 11$^\circ$ \citep{tho02gwlibv844herdiuma}. In combination with the previous works, our findings suggest the presence of vertically-expanded cool components at the outer region of the accretion disk during the appearance of the early superhumps, as suggested by \citet{mat09v455and} and \citet{nak13j0120}.


We find that the amplitudes of ordinary superhumps have no dependence on wavelength. \citet{pdot3} showed that amplitudes of superhumps are independent of inclination for not more than 80$^\circ$. On the other hand, systems with inclination higher than 80$^\circ$ show strong dependence on $i$ (see figure 90 of \cite{pdot3}). Based on these results, \citet{pdot3} suggest that the vertical structure in the superhump-light source is smaller than 10$^\circ$ in slope against the remaining disk. Taking \citet{pdot3} into consideration, our findings imply that the superhump light sources of HV Vir and J0120 are free of self-obscuration of the accretion disk. Because multicolor photometry of superhumps have been performed for only a few systems, we should collect further samples in order to better understand the light source of superhumps.

As illustrated in figure \ref{hvvir_Esh}, \ref{hvvir_sh}, and \ref{j0120_shvar}, we can notice the differences between the peaks of the light curves and colors in many panels. In early superhumps, the colors are at the bluest when the magnitudes are at the faintest. This trend is remarkably similar to that reported by \citet{mat09v455and} during the 2007 superoutburst of V455 And. \citet{mat09v455and} suggest that the temperature of the early superhump light source is lower than that of an underlying component. The present studies may further support the idea that the early superhump light source is a low-temperature, vertically expanded region at the outermost part of the accretion disk.

Regarding the ordinary superhumps, \citet{mat09v455and} reported that the bluest peak of the superhumps is prior to the maximum magnitude by phase of 0.15, and the reddest color corresponds to the bright minimum of the superhumps in phase. In contrast with \citet{mat09v455and}, such color variations were not found in our study. \citet{neu17j1222} recently published $B$, $V$, $R$, and $I$ photometry of the WZ Sge-type dwarf nova SSS J122221.7$-$311525 during the superoutburst, in which they studied color variations of the superhumps according to the stages. \citet{neu17j1222} point out that color variations are weak in stage A, whereas the bluest peaks correspond to the minimum magnitudes in stage B. As can be seen in figure \ref{hvvir_sh} and \ref{j0120_shvar}, some of our panels have a tendency that \citet{neu17j1222} noted on. In figure \ref{hvvir_sh} and \ref{j0120_shvar}, color variations are weak during stage A on February 20, December 10, and 11, despite the fact that the brightness amplitudes of stage A superhumps are the largest among the three stages \citep{pdot}. Also, the bluest peaks tend to coincide with the minimum magnitudes during stage B on February 26, 28, December 19, 22 and 23, and during stage A-B transition on February 21. In this respect, the color variations during the superhumps are in accordance with those reported by \citet{neu17j1222}.

In order to further examine the relation between superhump light curves and color variations, and compare our results with figure 9 of \citet{neu17j1222}, we made phase-averaged profiles of them in each stage, which are shown in figure \ref{shvarAB}. Here we define the amplitude ratio of color variations against superhump variations as $\Delta$.\footnote{For example, if the amplitude of color variation is 0.03 mag and the amplitude of superhumps is 0.2 mag, then $\Delta$ is 0.03/0.2 = 0.15.} We can roughly estimate $\Delta$ of HV Vir to be 0.2 for stage A, 0.4 for stage B, and 1 for stage C, respectively. As for J0120, we obtained 0.6 for stage A and 0.4 for stage B, respectively. Using figure 9 of \citet{neu17j1222}, we roughly estimated $\Delta$ of J122221.7$-$311525 to be 0.3 for stage A, 0.6 for the first half of stage B, 1 for the second half of stage B, and 1 for stage C, respectively. We summarize $\Delta$ of each object at different stages in table \ref{delta}. Although $\Delta$ for stage A of J0120 deviates from the other values, generally, $\Delta$ might be around 0.3 for stage A and larger than 0.4 for stage B. Our obtained values imply that stage B and C superhumps tend to show drastic color variations compared with stage A ones.

From the theoretical point, \citet{osa13v344lyrv1504cyg} proposed that the observed superhump period is determined mainly by three factors: dynamical procession, pressure effect, and wave-wave interaction. The wave-wave interaction is negligible compared with the other two factors. At the beginning of the appearance of ordinary superhumps, the pressure effect is immature so that this is negligible as well. As a result, the growing superhumps (stage A) can be regarded as the dynamical procession rate at the 3:1 resonance radius \citep{osa13v344lyrv1504cyg}. As the superoutburst proceeds, the pressure effect becomes dominant, which may be observed as stage B superhumps. Because the pressure effect leads to complicated physical processes in the accretion disk, we can expect that the color variations in stage B show diversity. On the other hand, a weak pressure effect causes the disk to show weak color variations, which may be observed as weak color variations in stage A superhumps. The observed color variations in WZ Sge-type dwarf novae to date seem to be consistent with our interpretation.

\citet{neu17j1222} investigated the post-superoutburst stage of J1222 and found that all the light curves are described by a broken power law. For example, $V$ band light curve in J1222 showed a break around $T$ ${\sim}$ 18, where $T$ is the day after the beginning of the rapid fading stage. \citet{neu17j1222} noted that the origin of a broken power-law decline is unclear, because of the lacking of papers both from the observational and theoretical points of view. In order to compare our results with those reported by \citet{neu17j1222} in J1222, we examined the post-superoutburst stage of HV Vir.\footnote{Due to the lack of the data of J0120 during the post-superoutburst stage, we only focused on HV Vir.} Figure \ref{post_gri} represents $g'$, $R_{\rm c}$, and $I_{\rm c}$ light curves during the post-superoutburst stage of HV Vir. Because of the apparent lack of the observations, we were unable to find a break in the light curves. Nevertheless, we calculated power-law indices of each band by fitting the light curves in units of flux. The obtained power-law indices were -0.25(4) in $g'$ band, -0.37(6) in $R_{\rm c}$ band, and -0.33(4) in $I_{\rm c}$ band, respectively. The obtained power-law index in $I_{\rm c}$ band of HV Vir is close to that of J1222 (-0.38(3), \citet{neu17j1222}). On the other hand, the power-law indices in $g'$ and $R_{\rm c}$ bands of HV Vir differ from those derived by \citet{neu17j1222} in J1222, who obtained -0.58(3) in $B$ band, -0.64(1) in $V$ band, and -0.54(2) in $R$ band, respectively. Overall, it should be noted that J1222 showed steep declines in $B$, $V$, and $R$ bands whereas HV Vir showed moderate declines in $g'$ and $R_{\rm c}$ bands. At presnt, it is unclear what causes such a difference between J1222 and HV Vir, although an apparent difference between them is that J1222 shows the type-E light curve and HV Vir shows the type-D one (\cite{kat13j1222}; \cite{kat15wzsge}). Further multicolor photometric studies are desirable in order to understand behavior of the post-superoutburst stage of WZ Sge-type dwarf novae.

\section{Summary}
We summarize our results as follows:
\begin{itemize}
\item{We detected early superhump periods of $P_{\rm Esh}$ =
  0.057093(45) d for HV Vir and $P_{\rm Esh}$ = 0.057147(15) d for J0120,
  respectively. The brightness minima of early superhumps correspond to the bluest peaks of color variations.}
\item{The amplitudes of early superhumps have dependence on
  wavelength. The amplitudes in the near-infrared range are larger than those in the optical range. $K_{\rm s}$ band shows the largest amplitudes while $g'$ band tends to show smaller amplitudes. This indicates that the early superhump light source is low-temperature and generated at the outer region of a
  vertically-extended accretion disk.}
\item{The amplitudes of ordinary superhumps seem to be independent of
  wavelength. This implies that the light source of ordinary
  superhumps are free of self-obscuration of the accretion disk. The superhump light source may be geometrically thin.}
\item{The bluest peaks of ordinary superhumps tend to coincide with the brightness minima. This tendency seems to be evident in stage B superhumps. On the other hand, color variations during stage A are weak. This may reflect significant pressure effects in stage B superhumps.}
\item{We obtained the power-law indices of HV Vir during the post-superoutburst stage and found that $g'$ and $R_{\rm c}$ bands showed moderate declining rates compared with $I_{\rm c}$ band. This behavior is in contrast with that observed in a candidate period bouncer SSS J122221.7$-$311525. The cause of the difference should be clarified by further multicolor photometry of WZ Sge-type dwarf novae during the post-superoutburst stage.}

\end{itemize}

\begin{ack}
We thank the anonymous referee for valuable comments on the manuscript of the paper. We acknowledge with thanks the variable star observations from the AAVSO and VSNET International Database contributed by observers worldwide and used in this research. 
\end{ack}

\bibliographystyle{pasjtest1}
\bibliography{cvs2016}

\begin{thebibliography}{}

\bibitem[{Ahn} et~al.(2012)]{sdss9}
  {Ahn}, C.~P., {et~al.}\ 2012, \apjs, 203, 21

\bibitem[{Araujo-Betancor} et~al.(2005)]{ara05v455and}
  {Araujo-Betancor}, S., {et~al.}\ 2005, \aap, 430, 629

\bibitem[Barwig et~al.(1992)]{bar92hvvir}
  Barwig, H., Mantel, K.~H., \& Ritter, H.\ 1992, \aap, 266, L5

\bibitem[{Drake} et~al.(2009)]{crts}
  {Drake}, A.~J., {et~al.}\ 2009, \apj, 696, 870

\bibitem[Hassall(1985)]{has85ektra}
  Hassall, B. J.~M.\ 1985, \mnras, 216, 335

\bibitem[Hellier(2001)]{hel01book}
  Hellier, C.\ 2001, Cataclysmic Variable Stars: How and why they vary (Berlin: Springer-Verlag)

\bibitem[{Hirose}, {Osaki}(1990)]{hir90SHexcess}
  {Hirose}, M., \& {Osaki}, Y.\ 1990, \pasj, 42, 135

\bibitem[Howell et~al.(1995)]{how95TOAD}
  Howell, S.~B., Szkody, P., \& Cannizzo, J.~K.\ 1995, \apj, 439, 337

\bibitem[{Imada} et~al.(2006)]{ima06tss0222}
  {Imada}, A., {Kubota}, K., {Kato}, T., {Nogami}, D., {Maehara}, H.,
  {Nakajima}, K., {Uemura}, M., \& {Ishioka}, R.\ 2006, \pasj, 58, L23

\bibitem[Ishioka et~al.(2001)]{ish01rzleo}
  Ishioka, R., {et~al.}\ 2001, \pasj, 53, 905

\bibitem[Ishioka et~al.(2003)]{ish03hvvir}
  Ishioka, R., {et~al.}\ 2003, \pasj, 55, 683

\bibitem[{Isogai} et~al.(2015)]{iso15ezlyn}
  {Isogai}, M., {Arai}, A., {Yonehara}, A., {Kawakita}, H., {Uemura}, M., \&
  {Nogami}, D.\ 2015, \pasj, 67, 7

\bibitem[Kato(2002)]{kat02wzsgeESH}
  Kato, T.\ 2002, \pasj, 54, L11

\bibitem[{Kato}(2015)]{kat15wzsge}
  {Kato}, T.\ 2015, \pasj, 67, 108

\bibitem[{Kato} et~al.(2013a)]{pdot4}
  {Kato}, T., {et~al.}\ 2013a, \pasj, 65, 23

\bibitem[{Kato} et~al.(2009)]{pdot}
  {Kato}, T., {et~al.}\ 2009, \pasj, 61, S395

\bibitem[{Kato} et~al.(2012a)]{pdot3}
  {Kato}, T., {et~al.}\ 2012a, \pasj, 64, 21

\bibitem[{Kato} et~al.(2012b)]{kat12DNSDSS}
  {Kato}, T., {Maehara}, H., \& {Uemura}, M.\ 2012b, \pasj, 64, 62

\bibitem[{Kato} et~al.(2013b)]{kat13j1222}
  {Kato}, T., {Monard}, B., {Hambsch}, F.-J., {Kiyota}, S., \& {Maehara}, H.\
  2013b, \pasj, 65, L11

\bibitem[Kato et~al.(2004a)]{kat04egcnc}
  Kato, T., Nogami, D., Matsumoto, K., \& Baba, H.\ 2004a, \pasj, 56, S109

\bibitem[{Kato}, {Osaki}(2013)]{kat13qfromstageA}
  {Kato}, T., \& {Osaki}, Y.\ 2013, \pasj, 65, 115

\bibitem[{Kato} et~al.(2001)]{kat01hvvir}
  {Kato}, T., {Sekine}, Y., \& {Hirata}, R.\ 2001, \pasj, 53, 1191

\bibitem[Kato et~al.(2004b)]{VSNET}
  Kato, T., Uemura, M., Ishioka, R., Nogami, D., Kunjaya, C., Baba, H., \&
  Yamaoka, H.\ 2004b, \pasj, 56, S1

\bibitem[Kato(2017)]{pdot9}
  Kato, T. et~al.\ 2017, \pasj, 69, 75

\bibitem[{Kotani} et~al.(2005)]{3me}
  {Kotani}, T., {et~al.}\ 2005, Nuovo Cimento C Geophysics Space Physics C, 28,
  755

\bibitem[{Lasota}(2001)]{las01DIDNXT}
  {Lasota}, J.-P.\ 2001, 45, 449

\bibitem[{Leibowitz} et~al.(1994)]{lei94hvvir}
  {Leibowitz}, E.~M., {Mendelson}, H., {Bruch}, A., {Duerbeck}, H.~W.,
  {Seitter}, W.~C., \& {Richter}, G.~A.\ 1994, \apj, 421, 771

\bibitem[{Lipunov} et~al.(2010)]{MASTER0}
  {Lipunov}, V., {et~al.}\ 2010, Advances in Astronomy, 2010, 349171

\bibitem[{Matsui} et~al.(2009)]{mat09v455and}
  {Matsui}, R., {et~al.}\ 2009, \pasj, 61, 1081

\bibitem[Meyer, Meyer-Hofmeister(1981)]{mey81DNoutburst}
  Meyer, F., \& Meyer-Hofmeister, E.\ 1981, \aap, 104, L10

\bibitem[{Nakagawa} et~al.(2013)]{nak13j0120}
  {Nakagawa}, S., {Noguchi}, R., {Iino}, E., {Ogura}, K., {Matsumoto}, K.,
  {Arai}, A., {Isogai}, M., \& {Uemura}, M.\ 2013, \pasj, 65, 70

\bibitem[{Neustroev} et~al.(2017)]{neu17j1222}
  {Neustroev}, V.~V., {et~al.}\ 2017, \mnras, 467, 597

\bibitem[{Nogami} et~al.(1997)]{nog97alcom}
  {Nogami}, D., {Kato}, T., {Baba}, H., {Matsumoto}, K., {Arimoto}, J.,
  {Tanabe}, K., \& {Ishikawa}, K.\ 1997, \apj, 490, 840

\bibitem[{Osaki}(1989)]{osa89suuma}
  {Osaki}, Y.\ 1989, \pasj, 41, 1005

\bibitem[{Osaki}(1996)]{osa96review}
  {Osaki}, Y.\ 1996, \pasp, 108, 39

\bibitem[{Osaki}, {Kato}(2013a)]{osa13v1504cygKepler}
  {Osaki}, Y., \& {Kato}, T.\ 2013a, \pasj, 65, 50

\bibitem[{Osaki}, {Kato}(2013b)]{osa13v344lyrv1504cyg}
  {Osaki}, Y., \& {Kato}, T.\ 2013b, \pasj, 65, 95

\bibitem[{Osaki}, {Meyer}(2002)]{osa02wzsgehump}
  {Osaki}, Y., \& {Meyer}, F.\ 2002, \aap, 383, 574

\bibitem[{Patterson}(2011)]{pat11CVdistance}
  {Patterson}, J.\ 2011, \mnras, 411, 2695

\bibitem[{Patterson} et~al.(1998)]{pat98egcnc}
  {Patterson}, J., {et~al.}\ 1998, \pasp, 110, 1290

\bibitem[Patterson et~al.(2002)]{pat02wzsge}
  Patterson, J., {et~al.}\ 2002, \pasp, 114, 721

\bibitem[Patterson et~al.(2003)]{pat03suumas}
  Patterson, J., {et~al.}\ 2003, \pasp, 115, 1308

\bibitem[{Ritter}, {Kolb}(2003)]{RKcat}
  {Ritter}, H., \& {Kolb}, U.\ 2003, \aap, 404, 301

\bibitem[Schoembs, Vogt(1980)]{sch80vwhyi}
  Schoembs, R., \& Vogt, N.\ 1980, \aap, 91, 25

\bibitem[{Shappee} et~al.(2014)]{ASASSN}
  {Shappee}, B.~J., {et~al.}\ 2014, \apj, 788, 48

\bibitem[{Smak}(1984)]{sma84DI}
  {Smak}, J.\ 1984, Acta, Astron, 34, 161

\bibitem[{Smith} et~al.(2002)]{2002AJ....123.2121S}
  {Smith}, J.~A., {et~al.}\ 2002, \aj, 123, 2121

\bibitem[Stellingwerf(1978)]{pdm}
  Stellingwerf, R.~F.\ 1978, \apj, 224, 953

\bibitem[Szkody et~al.(2002)]{szk02egcnchvvirHST}
  Szkody, P., {G\"ansicke}, B.~T., Sion, E.~M., \& Howell, S.~B.\ 2002, \apj,
  574, 950

\bibitem[{Thorstensen}(2003)]{tho03CVdistance}
  {Thorstensen}, J.~R.\ 2003, 126, 3017

\bibitem[Thorstensen et~al.(2002)]{tho02gwlibv844herdiuma}
  Thorstensen, J.~R., Patterson, J.~O., Kemp, J., \& Vennes, S.\ 2002, \pasp,
  114, 1108

\bibitem[Tovmassian et~al.(2003)]{tov03fsaur}
  Tovmassian, G., {et~al.}\ 2003, \pasp, 115, 725

\bibitem[{Tovmassian} et~al.(2007)]{2007ApJ...655..466T}
  {Tovmassian}, G.~H., {Zharikov}, S.~V., \& {Neustroev}, V.~V.\ 2007, \apj,
  655, 466

\bibitem[{Uemura} et~al.(2008)]{uem08j1021}
  {Uemura}, M., {et~al.}\ 2008, \pasj, 60, 227

\bibitem[{Uemura} et~al.(2012)]{uem12ESHrecon}
  {Uemura}, M., {Kato}, T., {Ohshima}, T., \& {Maehara}, H.\ 2012, \pasj, 64,
  92

\bibitem[Warner(1995)]{war95book}
  Warner, B.\ 1995, Cataclysmic Variable Stars (Cambridge: Cambridge University Press)

\bibitem[Whitehurst(1988)]{whi88tidal}
  Whitehurst, R.\ 1988, \mnras, 232, 35

\bibitem[Woudt, Warner(2002)]{wou02gwlib}
  Woudt, P.~A., \& Warner, B.\ 2002, 282, 433

\bibitem[{Woudt} et~al.(2012)]{wou12SDSSCRTSCVs}
  {Woudt}, P.~A., {Warner}, B., {de Bud{\'e}}, D., {Macfarlane}, S., {Schurch},
  M.~P.~E., \& {Zietsman}, E.\ 2012, \mnras, 421, 2414

\bibitem[{Yanagisawa} et~al.(2006)]{ISLE1}
  {Yanagisawa}, K., {et~al.}\ 2006, SPIE, 6269, 62693Q

\end{thebibliography}

\end{document}